\documentclass[aps,prx,twocolumn,superscriptaddress,citeautoscript,longbibliography]{revtex4-1}

\usepackage{amsmath,amssymb}
\usepackage{bm}
\usepackage{graphicx}
\usepackage{epstopdf}
\usepackage{latexsym}
\usepackage{subfigure}
\usepackage{color}
\usepackage{stmaryrd} 
\usepackage{hyperref} 

\definecolor{airforceblue}{rgb}{0.36, 0.54, 0.66}

\newcommand{\be}{\begin{equation}}
\newcommand{\ee}{\end{equation}}
\newcommand{\bea}{\begin{eqnarray}}
\newcommand{\eea}{\end{eqnarray}}

\def \nn{\nonumber}

\def \bJ{{\bf J}}
\def \br{{\bf r}}

\def \bq{{\bf q}}
\def \bk{{\bf k}}

\def \a{{\alpha}}

\def \d{{\delta}}
\def \w{{\omega}}
\def \s{{\sigma}}

\def \ve{{\varepsilon}}

\def \G{{\Gamma}}

\def \uar{{\uparrow}}
\def \dar{{\downarrow}}

\def \ba{\begin{align*}}
\def \ea{\end{align*}}

\newcounter{indice}

\def \mrm{\mathrm}

\def \mc{\mathcal}

\begin{document}

\title{Superconductivity in three-dimensional spin-orbit coupled semimetals}


 \author{Lucile Savary}
\email{savary@mit.edu}
 \affiliation{Department of Physics, Massachusetts Institute of
   Technology, 77 Massachusetts Ave., Cambridge, MA 02139}
 \affiliation{Laboratoire de physique, CNRS, \'{E}cole Normale Sup\'{e}rieure de Lyon, 46, all\'{e}e
   d'Italie, 69007 Lyon}
 \author{Jonathan Ruhman}
 \email{ruhman@mit.edu}
 \author{J\"{o}rn W. F. Venderbos}
 \author{Liang Fu}
\author{Patrick A. Lee}
 \affiliation{Department of Physics, Massachusetts Institute of
   Technology, 77 Massachusetts Ave., Cambridge, MA 02139}

\date{\today}
\begin{abstract}
Motivated by the experimental detection of superconductivity in the low-carrier density half-Heusler compound YPtBi, we study the pairing instabilities of three-dimensional strongly spin-orbit coupled semimetals with a quadratic band touching point. In these semimetals the electronic structure at the Fermi energy is described by spin $j=\frac32$ quasiparticles, which are fundamentally different from those in ordinary metals with spin $j=\frac12$.
We develop a general approach to analyzing pairing instabilities in $j=\frac32$ materials by decomposing the pair scattering interaction into irreducible channels, projecting them to the Fermi surface and deriving the corresponding Eliashberg theory.
Applying our method to a generic density-density interaction in YPtBi, we establish the following results:
{\em (i)} The pairing strength in the different symmetry channels uniquely encodes the $j=\frac32$ nature of the Fermi surface band structure---a manifestation of the fundamental difference with ordinary metals.
In particular, this implies that Anderson's theorem, which addresses the effect of spin-orbit coupling and disorder on pairing states of spin-$\frac12$ electrons, cannot be applied in this case.
{\em (ii)} The leading pairing instabilities are different for electron doping and hole doping. This originates from the different character of the electron and hole bands and implies that superconductivity depends on carrier type.
{\em (iii)} In the case of hole doping, which is relevant to YPtBi, we find two odd-parity pairing channels in close competition with $s$-wave pairing. One of these two channels is a multicomponent pairing channel, allowing for the possibility of time-reversal symmetry breaking.
{\em (iv)} In the case of Coulomb interactions mediated by the long-ranged electric polarization of the optical phonon modes, a significant coupling strength is generated in spite of the extremely low density of carriers. Furthermore, non-linear response and Fermi liquid corrections can favor non-$s$-wave pairing and potentially account for the experimentally observed $T_c$.
\end{abstract}

\maketitle

\section{Introduction}
\label{sec:introduction}

Increasingly many low density materials are being found to
superconduct. Examples include a rather diverse set of 2D and 3D
materials, doped topological insulators, semiconductors and
semimetals, such as Cu$_x$Bi$_2$Se$_3$ \cite{hor2010},
Pb$_{1-x}$Tl$_x$Te \cite{chernik1981}, single crystal
Bi \cite{Prakash2016}, Bi-based half-Heusler compounds---e.g.  YPtBi
and ErPdBi \cite{nakajima2015}, and of course doped SrTiO$_3$ has been
known to superconduct for more than 50 years \cite{schooley1964}. In
addition to a low density of carriers, many of these materials share a
number of other properties: sizeable spin-orbit coupling, pointers to
unconventional pairing, weak Coulomb repulsion due to a large dielectric screening, and in some cases ``proximity'' to a topological phase. In this context, questions which naturally arise are: What is the mechanism for such low-density superconductivity in those materials? Is it related to spin-orbit coupling? Is it particularly conducive to unconventional pairing?

Strong spin-orbit coupling causes the multiplicity of bands at high symmetry points in the Brillouin zone, such as the $\G$-point, to be larger than two, a signal that the bands themselves transform under a nontrivial/high-dimensional
representation of the crystal symmetry group. As a result, several of these materials host quasi-particles with large spin, e.g., $j=\frac32$ rather than the conventional $j=\frac12$. 
In particular, four-band $j=\frac32$ structures emerge from the $\Gamma_8$ states in cubic symmetry. They have been known for a long time \cite{luttinger1956,cardona1963fundamental,chadov2010}, but have recently attracted considerable interest due to their relevance to the {\em strongly-correlated} pyrochlore iridates where a flurry of unusual behaviors were uncovered \cite{yang2010,moon2013,savary2014,witczakkrempa2012,murray2015,yang2014}. Superconductivity, however, is absent in the iridates, where instead magnetic order develops at low
temperature \cite{matsuhira2011,witczakkrempa2014}.

In this context, the Bi-based half-Heusler superconductors such as
RPtBi and RPdBi, where R is either a rare-earth or Y/Lu, offer ideal
ground for the study of low-density superconductivity in spin-orbit
coupled systems and provide many potential examples of unconventional
superconductors \cite{pagliuso1999,goll2008,butch2011superconductivity,pan2013,tafti2013,xu2014,nikitin2015,nakajima2015,liu2016,zhang2016,kim2016}. Indeed, these materials share a very similar band structure with the paramagnetic pyrochlore iridates, but exhibit superconductivity at low temperature rather than magnetic order.  Most compounds in this family have a superconducting transition temperature close to 1~K, ranging from
approximately $0.7$~K in DyPdBi and $0.77$~K in YPtBi, to $1.6$~K in
YPdBi. The density of carriers (due to accidental doping) has been estimated at 10$^{18}$~cm$^{-3}$ in the Pt family, and is roughly 10$^{19}$~cm$^{-3}$ for the Pd materials \cite{nakajima2015}. The Fermi energy intercepts two bands with $j=\frac32$ character
\footnote{Though in Ref.~\cite{chadov2010} it is argued that the Fermi energy in  (Sm,Eu,Gd,Tb,Dy,Y)PdBi compounds might instead be close to another set, ``$\Gamma_6$''}
close to where they meet (at the $\Gamma$-point), and like in the pyrochlore iridates, ab initio calculations \cite{meinert2016,brydon2016,kim2016} and ARPES on YPtBi \cite{kim2016} show that {\em (i)} around the $\Gamma$-point two bands lie above the touching point while two bands lie below it {\em (ii)} pockets elsewhere in the Brillouin zone seem to be absent, at least in some of the compounds in the family (and hence the Fermi energy crosses only two bands). Most importantly, the predominantly Bi $p$-orbital character of the bands most likely produces only weak correlations, as is evident from a very large bandwidth, leaving only electronic and lattice (phonon) degrees of freedom as candidates for mediating superconductivity.

\begin{figure}
  \centering
  \includegraphics[width=\columnwidth]{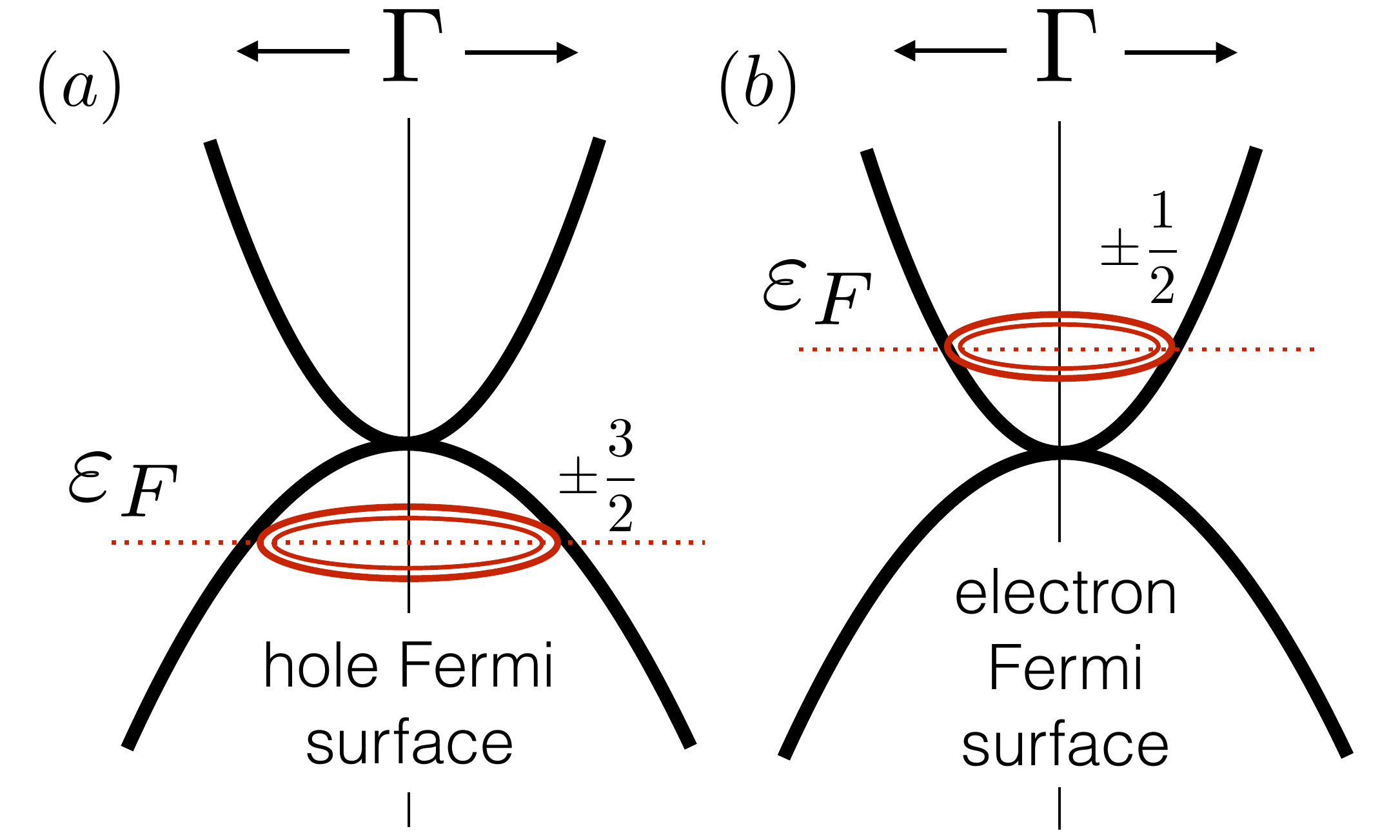}
  \caption{(a) Schematic electronic band structure of quadratic band
    crossing semimetals such as YPtBi. The touching of the $\Gamma_8$
    bands at is protected by symmetry. In the presence of strong
    spin-orbit coupling, i.e., coupling of the quasiparticle spin and
    crystal momentum, see Eqs.~\eqref{eq:ham} and Eqs.~\eqref{eq:36b},
    and with inversion symmetry the $\Gamma_8$ bands are split into
    two twofold degenerate bands away from $\Gamma$. Motivated by
    YPtBi, we assume that one of these bands curves upward, forming
    the electron band, and one curves downward, forming the hole
    band. In YPtBi, when the Fermi energy is in the hole band,
    corresponding to hole doping, the quasiparticle states on the
    Fermi surface are spin $\pm \frac32$ states, in the spherical approximation. (b) In the case of
    electron doping, which we also consider, the quasiparticle states on the Fermi surface are spin $\pm \frac12$ states.}
  \label{fig:bandstructure}
\end{figure}

Superconductivity at very low densities presents two challenges for conventional BCS theory: First, the Fermi energy can become so low that it is smaller than the relevant phonon energy, implying that the usual renormalization of the Coulomb repulsion from $\mu=\langle V_C\rangle_{\rm FS}$ (the Coulomb interaction strength averaged over the Fermi surface) to $\mu^*$ is no longer applicable, as is the case for doped SrTiO$_3$ \cite{ruhman2016}.
For the half-Heuslers, the Fermi energy is larger than the Debye frequency, but it is still of the same order~\footnote{This problem will be adressed in future work.}.
Second, in 3D, the density of states at the Fermi energy $N(0)$ goes
to zero as the carrier density is reduced. In standard BCS theory,
$T_c\propto \exp[-1/(N(0)V)]$, where $V$ is the pairing interaction
strength. For metals, the electron-phonon interaction is well-screened
so that $V$ is typically short ranged, and $T_c$ is expected to be
exponentially small as the density becomes small. This issue was
discussed a long time ago in a seminal work by Gurevich, Larkin and
Firsov (GLF)~\cite{gurevich1962}, who concluded that for a {\it
  short-ranged} attractive interaction 
superconductivity was not
expected at densities lower than $10^{19} \, \mrm{cm}^{-3}$, in line
with expectations. They proposed, however, that electron-phonon interactions could
circumvent the problem of a low density of states and efficiently
mediate superconductivity in ionic crystals where the lattice
distortion caused by an optical phonon generates polarization
(electric dipoles), and in turn effectively a {\em long-ranged} electron-phonon
interaction. Such an interaction is captured by the Fr\"olich
Hamiltonian \cite{frolich1954}, and, being long-ranged, benefits from
lower densities where 
it is not as effectively screened by other electrons in
the system. LDA calculations on YPtBi found the short-ranged electron-phonon $N(0)V$ to be
$0.02$ \cite{meinert2016}, much too small to support
superconductivity, but the numerical package used to obtain this
result did not capture the
Fr\"olich coupling \cite{Carla2015}, leaving open the possibility that the GLF
mechanism be responsible for superconductivity in this material. This is what we investigate in this manuscript.

To study superconductivity in a spin-orbit coupled multi-orbital system such as YPtBi, it is crucial to fully account for the $\Gamma_8$ character of the electronic states at the Fermi energy. This was addressed in an important recent paper by Brydon {\it et al.}\ \cite{brydon2016}, who pointed out that pairing of these spin $j=\frac32$ electrons was markedly different from pairing of ordinary $j=\frac12$ electrons: whereas in the latter case only spin-singlet and spin-triplet pairing states can be formed, since $\frac{1}{2}\otimes\frac{1}{2}=0\oplus1$, Cooper pairs composed of two $j=\frac32$ electrons can have higher spin, following $\frac{3}{2}\otimes\frac{3}{2}=0\oplus1\oplus2\oplus3$.

We will demonstrate that the non-trivial transformation of the bands also has important implications
for the pairing instabilities, and is most clearly seen when projecting
the pair scattering interaction onto the Fermi surface. Indeed, as
mentioned above, the symmetry group of the crystal enforces the
touching of all four bands at the Gamma point, but only requires
two-fold degeneracy away from it (by Kramers theorem through the
existence of inversion and time-reversal symmetries). Spin-orbit
coupling can then lead to a bending of the bands in opposite ways (see
Fig.~\ref{fig:bandstructure}), so that the Fermi energy crosses just two degenerate
bands with pseudospin index $\sigma$. Since only
electronic states close to the Fermi surface contribute to pairing,
the problem is superficially reminiscent of that with a
spin-$\frac{1}{2}$ degree of freedom. However, the structure of the Fermi
surface pseudospin states is very different, owing to the
$j=\frac{3}{2}$ nature of the $\Gamma_8$ bands. The projection of the
interactions onto the bands at the Fermi energy renders this fact
evident as the structure of the
spin-orbit coupled $\Gamma_8$ bands is reflected in the effective
coupling constants obtained from decomposing the projected interaction
into irreducible pairing channels, which themselves govern the
instabilities towards superconductivity.

This has deep implications for the pairing instabilities. 
For instance, we
will demonstrate that the effective coupling
constants of odd-parity pairing channels, which directly relate to
$T_c$, are different for the hole and electron Fermi surfaces, even
though their dispersions are similar. We will also explicitly show that
Anderson's theorem \cite{anderson1959} is inapplicable in the case of multiplets of bands
originating from spin-orbit coupling. 
Indeed, the conventional view on
the effect of spin-orbit coupling on
superconductivity is based on Anderson's Theorem \cite{anderson1959}:
it states that while spin is not a good quantum number, $T_c$ is not
affected if the $s$-wave singlet pairs are formed by Kramers
partners---which are guaranteed to exist by time-reversal symmetry. We
find that it
is not valid, however, when the bands originate from a larger
multiplet of states. In the present case, the effective coupling constants are affected
by the projection onto the Fermi surface, showing that the resulting effective single-band
description 
cannot be viewed as an ordinary
spin-$\frac12$ system, and rendering Anderson's Theorem
inapplicable. 
Both the hole/electron band dependence of $T_c$ and the breakdown of
Anderson's theorem are clear manifestations of the
key significance of spin-orbit coupling of $j=\frac32$ bands for
superconducting instabilities.

In this paper we develop a general approach to studying pairing
instabilities in doped spin-orbit coupled $j=\frac32$ systems with
quadratic band touching dispersion. We identify the relevant symmetry
quantum numbers and decompose the pair scattering interaction into
irreducible pairing channels. This decomposition reveals the natural
mean-field decouplings, which can be used to derive the corresponding
BCS (or Eliashberg) gap equations. Our approach is independent of the
symmetry group of the normal state, though we apply the formalism to the
half-Heusler material YPtBi and for ease of presentation generally assume full spherical symmetry
before discussing the effect of cubic crystal fields.

The remainder of the manuscript is organized as follows. We first
provide the band structure model relevant to the half-Heuslers, and
introduce the density-density interaction we will be considering
throughout. We then turn to a rewriting of the interaction into
irreducible representation components, and consider the projection of
these terms onto the valence bands. We then derive the appropriate
Eliashberg equations before moving on to a discussion of the results.



\section{Band structure and interactions}
\label{sec:band-struct-inter}

We start our analysis by introducing the model appropriate to describe the low-energy electronic physics of nonmagnetic half-Heuslers. The electronic action consists of two terms: a quadratic term, representing the free kinetic part, and a quartic term describing the interactions. We write
\begin{equation}
  \label{eq:1}
  \mathcal{S}=\mc S_0+\mathcal{S}_{\rm
  int}.
\end{equation}
In what follows we discuss each of these terms in detail.

\subsection{Band Hamiltonian}

The free quadratic part of the action is given by
\be
\mc S_0 = \sum_\mathbf{r}\int d\tau
  \psi_{\mathbf{r}\tau}^\dagger\big[\partial_\tau+\mathcal{H}_0(-i\nabla)\big]\psi_{\mathbf{r}\tau}
\ee
where $\psi^\dag = (\psi_{\frac32}^\dag,\psi_{\frac12}^\dag,\psi_{-\frac12}^\dag,\psi_{-\frac32}^\dag)$ is a four-component creation operator of spin $j=\frac32$ fermions and
 \cite{dresselhaus1955}
\begin{multline}
\mathcal{H}_0(\mathbf{k})=\alpha_1 \mathbf{k}^2 + \alpha_2  \left(\mathbf{k} \cdot \mathbf{J}\right)^2 + \alpha_3  \left(k_x^2 J_x^2 + k_y^2 J_y^2 +k_z^2 J_z^2\right)  \\
+ \alpha_4\mathbf{k}\cdot\mathbf{T}-\mu  \label{eq:ham}
\end{multline}
In the first line, $\bJ = (J_{x}, J_y,J_z)$ are the three $4\times4$
spin matrices of $j=\frac32$ electrons. In addition, $\mu$ is the
chemical potential (such that $\mu>0$, resp.\ $\mu<0$, corresponds to
electron, resp.\ hole doping) and $\alpha_{1,2,3,4}$ are material-dependent parameters characterizing the electronic band structure. When $\alpha_3=\alpha_4=0$ the system has full spherical symmetry. The term proportional to $\alpha_3$ reduces the symmetry to cubic crystal symmetry while $\mathbf{k}\cdot\mathbf{T}$, with $T_x=\{J_x,J_y^2-J_z^2\}$ and $T_{y,z}$ given by cyclic permutations, is only allowed in an inversion symmetry broken tetrahedral crystal field.

The Hamiltonian of Eq.~\eqref{eq:ham} can be usefully rewritten in terms of anticommuting $\Gamma$-matrices (given in the Supplementary Material) and the five $d$-wave functions $d_a(\mathbf{k})$ quadratic in momentum. One obtains
\begin{multline}
\mathcal{H}_0(\mathbf{k})=c_0\mathbf{k}^2+c_1\sum_{a=1}^3 d_a(\mathbf{k})\Gamma_a+c_2\sum_{a=4}^5 d_a(\mathbf{k})\Gamma_a \\+c_3\mathbf{k}\cdot\mathbf{T}-\mu.
\label{eq:36b}
\end{multline}
The coefficient $c_0$ measures the particle-hole asymmetry in the band structure, while
$|c_{1}-c_2|$ measures its cubic anisotropy: $c_1=c_2$ corresponds to
full spherical symmetry, whereas $c_1 \neq c_2$ implies a splitting of
the five $d$-wave into $T_{2g}$ and $E_{g}$ subsets. When $c_3=0$, the
system has both time-reversal and inversion symmetry, mandating a
two-fold degeneracy at each momentum $\bk$. In that case, a simple expression for the energy eigenvalues can be obtained and is given by $\mathsf{E}_\nu(\mathbf{k})=c_0\mathbf{k}^2+\nu
E_\bk-\mu$, where $\nu$ denotes the band and corresponds to $+1$ for
electron bands and $-1$ for the hole ones, and $E_\bk=(c_1^2\sum_{a=1}^3d_a^2+c_2^2\sum_{a=4}^5d_a^2 )^{1/2}$. The Kramers degeneracy is labeled by the index $\s$. Thus, overall the Bloch states are denoted by
$|{\bf{k}} ,\nu, \s \rangle$.

The condition $|c_0|\leq |c_1|/\sqrt{6}$ guarantees that two bands always curve upwards, forming the conduction band, while the other two curve downwards (see
Fig.~\ref{fig:bandstructure}).
In the presence of full spherical symmetry (i.e.,
$\alpha_3=\alpha_4=0$, or $c_1=c_2$ and $c_3=0$),
$\mathbf{k}\cdot\mathbf{J}$ commutes with the Hamiltonian, as may be
seen directly from Eq.~\eqref{eq:ham}, and so the projection of the
spin along $\mathbf{\hat{k}}$ is a good quantum number. In other words, the
quantization axis of the spin is locked to $\bf k$. The hole and
electron bands in that case may be labeled according to $3/2$ or
$1/2$, following $\nu\, {\rm sign}(c_1)$: the band with $\nu\,{\rm
  sign}c_1=1$ is the $3/2$ band while that with $\nu\,{\rm
  sign}c_1=-1$ is $1/2$.
For the parameters of YPtBi \cite{brydon2016} the hole $\nu=-1$
(electron $\nu=+1$) band is the
pair with a projected moment of $\pm {3 \over 2}$ ($1\over 2$). 
It is important to note that even though the electron and hole band appear to have a similar dispersion (i.e., both look like quadratic bands, curving downward and upward, respectively), their structure, as encoded in the eigenstates, is inherently different. For example, one has $\langle{\bf k},\frac32,\s|\tilde{J}_\mathbf{k}^+|{\bf k},\frac32,-\s\rangle=0$, while $\langle{\bf k},\frac12,\s|\tilde{J}_\mathbf{k}^+|{\bf k},\frac12,-\s\rangle\neq0$, where $\tilde{J}_\mathbf{k}^\pm$ are the raising and lowering operators corresponding to $\tilde{J}_\mathbf{k}^z=\mathbf{J}\cdot\mathbf{\hat{k}}$.

While the half-Heusler compounds---space group
F$\overline{4}$$3$m---actually lack inversion symmetry, ab initio calculations
suggest that inversion breaking has only a weak effect on the band
structure \cite{chadov2010} as compared, e.g., to the cubic
Fd$\overline{3}$m in the pyrochlore iridates. Since the consequences of spin-orbit coupling seem to be most important, 
we expect that many of the notable results we derive hold in a similar
form in the absence of inversion symmetry. Therefore, in the bulk of
the manuscript we neglect the effects of the absence of inversion
symmetry---namely we set $c_3=0$, $\alpha_4=0$, eliminating the terms
linear in $\mathbf{k}$ in the band Hamiltonian. This allows us to
carry out analytical calculations which in turn help provide a deeper
understanding of the problem, and are also directly relevant to cubic
materials with inversion symmetry. 

\subsection{Interactions}

As explained, we focus here on the attractive interaction mediated by optical phonons through the Fr\"olich electron-phonon coupling. The interaction term in Eq. \eqref{eq:ham} is part of the long-ranged Coulomb (density-density) interaction. Collecting position and imaginary time variables in the the index $x=(\br,\tau)$, the interaction takes the form
\be
\mathcal{S}_{\text{int}}= \frac12 \int_{x,x'}  V(x-x') \psi^\dagger_{x} \psi_{x} \psi^\dagger_{x'} \psi_{x'} ,   \label{eq:2-1}
\ee
where $\int_x = \sum_\br \int d\tau$ and the interaction $V=
V(\mathbf{r},\tau)$ has Fourier and Matsubara components
\be
V(\bq,\w) = {4\pi e^2 \over \ve(\bq,\w)q^2}\,. \label{Coulomb}
\ee
The total dielectric function has three contributions
 \be
\ve(\bq,\w) = \ve_{\infty} +\ve_c(\bq,\w)+\ve_e(\bq,\w).\label{epsilon}
\ee
$\ve_\infty$ comes from interband transitions, and
\be
 \ve_c(\bq,\w) = {\ve_0 - \ve_\infty \over 1 + [\w/\w_T(\bq)]^2}\,
\label{epsc}
\ee
is the polarization in {\em Matsubara} frequency due to a polar phonon mode.
Note that for simplicity, we have considered the case of a single phonon mode. $\w_T$ is the frequency of the transverse optical mode, which is related to the longitudinal one through the Lyddane-Sachs-Teller (LST) relation $\w_L = \sqrt{\ve_0 / \ve_\infty }\,\w_T$. Finally, the last term, $\varepsilon_e$, is the electronic polarization, taken within the random-phase-approximation (RPA) to be
\be
\ve_e(\bq,\w) = -{4\pi e^2 \over q^2} \Pi_e(\bq,\w)\,,
\ee
where $\Pi_e$ is the electronic polarization, which we will later take
in the Thomas-Fermi approximation, where
$4\pi e^2\Pi_e$ is replaced by $-q_{\rm TF}^2$. We leave the study
of the full polarization function, which includes the non-Fermi
liquid $V(q)\propto 1/q$ regime of the undoped quadratic band touching
system \cite{moon2013,boettcher2016}, to a future publication.

\section{Irreducible pairing channels}
\label{sec:irred-repr-part}

The next step in our analysis is to obtain and classify the set of
irreducible pairing channels. To this end, we rewrite the
density-density interaction Eq.~\eqref{eq:2-1} as a pair scattering
Hamiltonian and then decompose the pair scattering terms into
irreducible scattering vertices. Pairing channels are labeled by the
quantum numbers of the Cooper pairs, and this labeling applies to
irreducible scattering vertices as well. The symmetry quantum numbers
of the Cooper pairs clearly depend on the symmetry group of the
system. In the presence of full rotational symmetry (i.e., ignoring
cubic anistropy), the Cooper pair quantum numbers are given by its
``spin'' angular momentum $S$, which corresponds to the band index, its
orbital angular momentum $L$, which corresponds to the momentum
dependence of the pairing function, and its
total angular momentum $J=L+S$ (not be confused with the spin operators $J_{x,y,z}$). For ease of presentation and clarity we will
present all derivations in the language of spherical symmetry, and
later indicate what the modifications are in lower symmetry.

In the present case of $j=\frac32$ fermions, the spin angular momentum of the Cooper pair can take the values $S=0,1,2,3$ \cite{fang2015,yang2016}. It is instructive to compare this to the more familiar case of spin $j=\frac12$ fermions, which can form Cooper pairs of $S=0$ (singlet) or $S=1$ (triplet). In this case, a two-body density-density interaction can be decomposed into singlet and triplet scattering vertices. More precisely, if $c_{\mathbf{k}}^\dagger = (c_{\mathbf{k}\uar}^\dagger,c_{\mathbf{k}\dar}^\dagger)$ are the creation operators of spin-$\frac12$ fermions, then one has the identity
\begin{multline}
(c_{\bk}^\dagger c_{\bk'})(c_{-\bk}^\dagger   c_{-\bk'}) =\frac{1}{2} [c_{\bk}^\dagger i\sigma^y (c_{-\bk}^\dagger)^T][(i\sigma^yc_{-\bk'})^Tc_{\bk'}] \\
+\frac{1}{2}[c_{\bk}^\dagger {\boldsymbol{\sigma}}  i\sigma^y (c_{-\bk}^\dagger)^T] \cdot [(i\sigma^yc_{-\bk'})^T{\boldsymbol{\sigma}} c_{\bk'}],  \label{eq:30}
\end{multline}
where the dot product is between components of ${\boldsymbol{\sigma}}$, i.e.\ $[c_{\bk}^\dagger {\boldsymbol{\sigma}}  i\sigma^y (c_{-\bk}^\dagger)^T] \cdot [(i\sigma^yc_{-\bk'})^T{\boldsymbol{\sigma}} c_{\bk'}]\equiv \sum_\alpha[c_{\bk}^\dagger \sigma^\alpha  i\sigma^y (c_{-\bk}^\dagger)^T][(i\sigma^yc_{-\bk'})^T\sigma^\alpha c_{\bk'}]$. The appearance of $i\sigma^y$ guarantees the symmetry and antisymmetry
of the spin part of the Cooper pair wave function for triplet and
singlet pairing because 
$i\sigma^y$ relates the fundamental and adjoint representations of SU(2), such that $i\sigma^y (c^\dagger_{-\bf k})^T$ transforms as $c_{\bk}$. Note that $i\sigma^y$ is antisymmetric, $(i\sigma^y)^T=-i\sigma^y$, and together with conjugation acts as a time-reversal operation on spin: $(-i\sigma^y) \boldsymbol{\sigma}^* i\sigma^y = -\boldsymbol{\sigma}$.

%

In a manner fully analogous to Eq.~\eqref{eq:30}, the interaction of Eq.~\eqref{eq:2-1}, which describes two-body density-density interactions of spin-$\frac32$ fermions, can be decomposed into irreducible spin channels labeled by $S=0,1,2,3$. In this decomposition we make use of the antisymmetric matrix $\gamma$, which serves as the analog of $i\sigma^y$. In particular, $\gamma$ satisfies $\gamma^T=  -\gamma$ and $\gamma^T \bJ^* \gamma =   -\bJ$, and it relates the fundamental and adjoint representations of spin-$\frac32$ fermions: $\gamma  (\psi^\dagger)^T$ transforms as $\psi$ under rotations. In the usual basis of $3/2$ eigenstates, the matrix $\gamma$ is given explicitly by
\be
\gamma = \begin{pmatrix} 0  & 0 & 0 & 1 \\ 0& 0& -1 &0 \\ 0& 1& 0& 0\\ -1& 0& 0& 0 \end{pmatrix}.
\ee
Now, taking the Fourier transform of Eq.~\eqref{eq:2-1} and going to Matsubara frequency space, the density-density product of operators can be decomposed into pair scattering terms as (suppressing the frequency index of the operators)
\begin{multline}
(\psi^\dagger_{\bk} \psi_{\bk'})(\psi^\dagger_{-\bk} \psi_{-\bk'}) =  \\
\frac14 \sum_{S}
     \psi^\dagger_{\bk}\vec{M}_{S}\gamma (\psi^\dagger_{-\bk})^T\cdot(\psi_{-\bk'})^T\gamma^T\vec{M}^\dagger_{S}\psi_{\bk'}, \label{eq:31}
\end{multline}
where the sum is over irreducible spin channels $S=0,1,2,3$. The
matrices $\vec{M}_{S}$ are $4\times 4$ matrices such that
$\psi^\dagger_{\bk}M^\alpha_{S}\gamma (\psi^\dagger_{-\bk})^T$ creates
a Cooper pair with total spin $S$. There are $2S+1$ matrices collected
in the vector $\vec{M}_{S}$, corresponding to the degeneracy of the
channel $S$. The matrices $\vec{M}_{S}$ are normalized such that each
component of the vector, $M^\alpha_{S}$, satisfies ${\rm  Tr}[ M^\alpha_S
(M^\alpha_S)^\dagger]=4$ (no implicit summation over $\alpha$). They
are listed in \ref{tab:Mmatrices-cubic-main}. For instance, in case of
$S=0$ the single matrix $M_{S=0}$ is simply equal to the identity; for
$S=1$ one has $\vec{M}_{S=1} \propto (J_x,J_y,J_z)$. We note in
passing that since the $S=0$ and $S=1$ channels are commonly referred
to as spin-singlet and spin-triplet, the $S=2$ and $S=3$ channels are
sometimes referred to as spin-quintet and spin-septet (e.g., in
Refs.~\onlinecite{brydon2016} and \onlinecite{kim2016}). 
For the $L>0$ channels this is unrelated to the actual multiplicity of
the Cooper pair pairing channels, which is determined by $J$ (and not
$S$) and equal to $2J+1$.

We note that the decomposition Eq.~\eqref{eq:31} can be viewed as a
Fierz identity, as can Eq.~\eqref{eq:30} (see Supp.\ Mat.). Furthermore, at this stage it is worth pointing out that the $S=0$ channels in Eq.~\eqref{eq:31} and Eq.~\eqref{eq:30}, which are associated with $s$-wave pairing, have different numerical prefactors: $1/4$ and $1/2$, respectively. In fact, the numerical prefactors in Eqs.~\eqref{eq:31} and Eq.~\eqref{eq:30} are equal to $1/(2j+1)$ and simply follow from the Fierz identities. Below we will find that these prefactors are important for the effective coupling constants in the $s$-wave channel.

\begin{table}
  \centering
  \begin{tabular}[c]{lclc}
\hline\hline
$S$ & Even/odd & $R$ &  $\vec{M}_R$  \\
\hline
$0$ & Even & $A_{1g}$ &  $I_4$  \\
$2$ & Even & $E_g$ &  $(\Gamma_4,\Gamma_5)$  \\
 &  & $ T_{2g}$ &  $(\Gamma_1,\Gamma_2,\Gamma_3)$ \\
\hline
$1$ & Odd & $T_{1g}$ &  $\frac{2}{\sqrt{5}}(J_x,J_y,J_z)$  \\
$3$ & Odd & $A_{2g}$ &  $\frac{2}{\sqrt{3}}(J_xJ_yJ_z+J_zJ_yJ_x)$  \\
 &&$T_{1g}$ & $\frac{-41}{6\sqrt{5}}(J_x,J_y,J_z)+\frac{2\sqrt{5}}{3}(J_x^3,J_y^3,J_z^3)$  \\
&& $T_{2g}$ &  $\frac{1}{\sqrt{3}}(T_x,T_y,T_z)$  \\
\hline\hline
  \end{tabular}
  \caption{List of spin pairing matrices $\vec M_S$ introduced in Eq. \eqref{eq:31}. Quasiparticles with $j=\frac32$ can form Cooper pairs with spin $S=0,1,2,3$; a Cooper pair of spin $S$ is created by the operator $\psi^\dagger_{\bk}\vec{M}_{S}\gamma (\psi^\dagger_{-\bk})^T$. Fermi statistics requires that the overall pairing function is even. Thus, the pairing matrices with $S$ even are allowed locally (i.e. momentum independent). On the other hand the odd matrices here must be further multiplied by an odd power of momentum, which leads to a richer classification. In Table ~\ref{tab:Nmatrices-cubic-main} we present the resulting representations in the case of a a single power of momentum. Note that in cubic symmetry the SO(3) representations labeled by $S$ are split into cubic representations labeled by $R$. }
  \label{tab:Mmatrices-cubic-main}
\end{table}

\begin{table}[htbp]
  \centering
  \begin{tabular}[c]{cc}
    \hline\hline
    $R'$ & $\vec{N}_{R'}(\mathbf{k})$ \\
    \hline
    $A_{1u}$ & $\frac{2}{\sqrt{5}}\mathbf{k}\cdot\mathbf{J}$ \\
    $T_{1u}$ & $\frac{\sqrt{6}}{\sqrt{5}}\mathbf{J}\times\mathbf{k}$ \\
    $E_u$ & $\frac{\sqrt{2}}{\sqrt{5}}(-J_x k_x-J_y k_y+2J_z
            k_z,\sqrt{3}(J_x k_x-J_y k_y))$ \\
    $T_{2u}$ & $\frac{\sqrt{6}}{\sqrt{5}}(J_y k_z+J_z k_y,J_x k_z+J_zk_x,J_x k_y+J_y k_x)$ \\
    \hline
    $T_{2u}$ & $-2\Gamma_{45}\mathbf{k}$ \\
    $A_{1u}$ &
               ${\boldsymbol{\mathcal{J}}}\cdot\mathbf{k}$ \\
    $E_u$ &
            $\frac{1}{\sqrt{2}}(-k_x\mathcal{J}_x-k_y\mathcal{J}_y+2k_z\mathcal{J}_z,\sqrt{3}(k_x\mathcal{J}_x-k_y\mathcal{J}_y))$ \\
    $T_{1u}$ &
               $\frac{\sqrt{3}}{\sqrt{2}}{\boldsymbol{\mathcal{J}}}\times\mathbf{k}$
                                \\
    $T_{2u}$ & $\frac{\sqrt{3}}{\sqrt{2}}(\mathcal{J}_y
               k_z+\mathcal{J}_z k_y,\mathcal{J}_x k_z+\mathcal{J}_z
               k_x,\mathcal{J}_x k_y+\mathcal{J}_y k_x)$ \\
    $A_{2u}$ & $\frac{1}{\sqrt{3}}\mathbf{T}\cdot\mathbf{k}$ \\
    $E_u$ & $\frac{1}{\sqrt{6}}(T_x k_x+T_y k_y-2T_z
            k_z,\sqrt{3}(T_x k_x-T_y k_y))$ \\
    $T_{1u}$ & $\frac{1}{\sqrt{2}}(T_y k_z+T_z k_y,T_x k_z+T_z k_x,T_x
               k_y+T_y k_x)$ \\
    $T_{2u}$ & $\frac{1}{\sqrt{2}}\mathbf{T}\times\mathbf{k}$ \\
    \hline\hline
  \end{tabular}
  \caption{List of odd-parity total angular momentum pairing matrices
    $N_{J}(\mathbf{k})$ in cubic
    symmetry with inversion $O_h$ constructed from the odd matrices in
    Table~\ref{tab:Mmatrices-cubic-main} and a factor of $k^\mu$. A Cooper pair with total angular
    momentum $J$ is created by one of the operators
    $\psi^\dagger_\mathbf{k}\vec{N}_J(\mathbf{k})\gamma
    (\psi_{-\mathbf{k}}^\dagger)^T$. We defined
    ${\boldsymbol{\mathcal{J}}}=\frac{-41}{6\sqrt{5}}\mathbf{J}+\frac{2\sqrt{5}}{3}(J_x^3,J_y^3,J_z^3)$. The
  horizontal line separates $S=1$ ($T_{1g}$) from $S=3$
  ($A_{1g}+T_{1g}+T_{2g}$) channels.}
  \label{tab:Nmatrices-cubic-main}
\end{table}

We have now arrived at an expression for the interaction
Eq.~\eqref{eq:2-1} of the following form, considering only zero linear
momentum Cooper pairs
\begin{multline}
\label{eq:SintMS}
\mathcal{S}_{\rm
     int}=\frac{1}{8\beta\mc V}\sum_{k,k'} V(k-k') \\
  \times    \sum_{S} \psi^\dagger_{k}\vec{M}_{S}\gamma (\psi^\dagger_{-k})^T\cdot(\psi_{-k'})^T\gamma^T\vec{M}^\dagger_{S}\psi_{k'},
\end{multline}
where $\mc V$ is the total volume, $\beta$ is the inverse temperature $\beta=1/(k_{\rm B}T)$, and we have collected the momentum $\bk$ and fermionic Matsubara frequencies $\w$ in $k= (\bk,\w)$.

To proceed with the derivation of irreducible pairing channels, we now
focus on the orbital angular momentum of the Cooper pairs. The orbital
angular momentum can be labeled by the quantum numbers $L$ and
$\mathcal{M}_L$, where $\mathcal{M}_L$ is the familiar $(2L+1)$-fold
degenerate magnetic quantum number, and the orbital part of the Cooper
pair wave function is given by the spherical harmonics
$Y_{L\mathcal{M}_L}(\hat \bk)$. Fermi statistics requires that $L$ is
even (odd) when $S$ is even (odd). 
The irreducible pairing channels are classified by the total angular momentum $J=L+S$ of the Cooper pairs. Using the rules of composition of angular momentum, we take the spherical harmonics $Y_{L\mathcal{M}_L}(\hat \bk)$ and spin matrices $\vec M_S$, and construct the spin-orbit coupled matrices $\vec{N}_{J}(\hat \bk)$ such that $\psi^\dagger_{\bk}N^\alpha_{J}(\mathbf{\hat{k}})\gamma (\psi^\dagger_{-\bk})^T$ creates a Cooper pair with total angular momentum $J$. The dimension of the vector $\vec N_J$ is $2J+1$ and can be labeled by the index $\mathcal{M}_J$.

Let us take the case $L=1$ as an example. Then, Fermi statistics
restricts $S$ to be odd: $S=1,3$. The combination $(L,S)=(1,1)$ gives
rise to the multiplets $J=0,1,2$; from $(L,S)=(1,3)$ one finds
$J=2,3,4$. Then, using the $p$-wave spherical harmonics
$Y_{1\mathcal{M}_1}(\hat \bk)\sim \hat \bk$ and the odd channels of
the pair scattering interaction of Eq.~\eqref{eq:31}, we obtain the irreducible pair scattering vertices labeled by $J$ as
\begin{multline}
\label{eq:4}
\mathbf{\hat{k}}\cdot\mathbf{\hat{k}}'\sum_{S=1,3} \psi^\dagger_{k}\vec{M}_{S}\gamma (\psi^\dagger_{-k})^T\cdot(\psi_{-k'})^T\gamma^T\vec{M}^\dagger_{S}\psi_{k'} \\
 =\frac{1}{3}\sum_{J}
\psi^\dagger_{k}\vec{N}_{J}(\mathbf{\hat{k}})\gamma
(\psi^\dagger_{-k})^T\cdot(\psi_{-k'})^T\gamma^T\vec{N}_{J}^\dagger(\mathbf{\hat{k}})\psi_{k'},
\end{multline}
where the sum over $J$ is here a short-hand notation for a sum over
the odd-$S$ combinations $(L,S)=(1,1)$ ($J=0,1,2$) and $(L,S)=(1,3)$ ($J=2,3,4$), and the matrices $\vec{N}_{J}(\mathbf{\hat{k}})$ are normalized according to $\frac{1}{4\pi}\int d\mathbf{\hat{k}}{\rm
  Tr}[N^\alpha_{J}(\mathbf{\hat{k}})N^\alpha_{J}{}^\dagger(\mathbf{\hat{k}})]=4$
(no implicit $\alpha$ summation). We list the matrices
$\vec{N}_{J}(\mathbf{\hat{k}})$ in cubic representations in Table~\ref{tab:Nmatrices-cubic-main}. Note that $\sum_{\mathcal{M}_1} Y^*_{1\mathcal{M}_1}(\hat\bk)Y_{1\mathcal{M}_1}(\hat\bk') = 3\hat \bk \cdot \hat  \bk'/4\pi$.

Equation~\eqref{eq:4} allows us to fully decompose the density-density interaction $V(\bq,\w)$ into irreducible pairing vertices. In the presence of full rotational symmetry, the interaction can be expanded as a sum over products of spherical harmonics. Here and in the remainder of this paper we shall restrict the expansion to linear $p$-wave order in $\bk$, i.e., to the order $L=1$, and write
\begin{multline}
V(\mathbf{k}-\mathbf{k}',\w-\w')=  V_0(\w-\w')+ \\
3V_1(\w-\w')\mathbf{k}\cdot\mathbf{k}'+\cdots   \label{eq:38}
\end{multline}
(see Supp.\ Mat.) Note that the interaction parameters $V_{0,1,\ldots}$ can still depend
on the magnitude of $\bk,\bk'$; this is suppressed as it does not
affect the rest of the analysis (and later we will take
$|\mathbf{k}|=|\mathbf{k}'|=k_{\rm F}$). We then arrive at the final form of the interaction term given by
\begin{multline}
\mathcal{S}_{\rm int} = \frac{1}{8\beta \mc V}\sum_{k,k'}\sum_J \hat{V}^{J}_{\alpha\beta\gamma\delta}(\bk,\bk';\w-\w')  \\
 \times\psi_{k \alpha}^\dagger   \psi_{-k \beta }^\dagger \psi_{-k' \gamma }\psi_{ k' \delta},   \label{eq:39}
\end{multline}
where here the sum over $J$ runs over both the even and odd
representations, i.e.\ the
combinations $(L,S)=(0,0)$ ($J=0$), $(L,S)=(0,2)$ ($J=2$), and
$(L,S)=(1,1)$ ($J=0,1,2$) and $(L,S)=(1,3)$ ($J=2,3,4$), and the spin-dependent pair scattering vertices $\hat{V}^{J}_{\alpha\beta\gamma\delta}$ take the form
\begin{equation}
  \label{eq:40}
\hat{V}_{\alpha\beta\gamma\delta}^{J} =\begin{cases}
V_0[\vec{M}_{J}\gamma]_{\alpha\beta}\cdot[\gamma^T \vec{M}^\dagger_{J}]_{\gamma\delta}   & \mbox{for }S=even\\
V_1[\vec{N}_{J}(\bk)\gamma]_{\alpha\beta}\cdot[\gamma^T \vec{N}^\dagger_{J}(\bk')]_{\gamma\delta}  &\mbox{for }S=odd
\end{cases}.
\end{equation}
Here we have used that $\vec{N}_{J}(\mathbf{k}) = \vec{M}_{J}$ whenever $S$ is even, since $L=0$ in this case.

Up to this point in this section, we have particularized to the case
of full spherical symmetry, which allowed us to label the irreducible
pairing channels by symmetry quantum number $J$. In a cubic crystal,
however, pairing channels are labeled by the representations of the
cubic point group. Importantly, the decomposition schemes of
Eqs.~\eqref{eq:SintMS} and \eqref{eq:39} remain valid (because
Eqs.~(\ref{eq:31},\ref{eq:4},\ref{eq:38}) do), but the sums over
the symmetry quantum numbers $S$, $L$, and $J$, all of which are
labels of SO(3) representations, must be replaced by sums over cubic
representations $R$. The effect of lower symmetry, i.e., cubic instead
of full spherical symmetry, is to lift some of the degeneracies of the $J>1$ channels. For instance, in a cubic environment the even-parity $L=0$ channels acquire the symmetry labels
\begin{eqnarray}
J=0 &\rightarrow &  A_{1g} , \nonumber \\
J=2 &\rightarrow &  E_g+T_{2g} ,   \label{eq:cubiceven}
\end{eqnarray}
whereas the odd-parity pairing channels become
\begin{eqnarray}
J=0 &\rightarrow &  A_{1u} ,\nonumber  \\
J=1 &\rightarrow &  T_{1u} ,\nonumber  \\
J=2 &\rightarrow &  E_u+T_{2u} ,\nonumber  \\
J=3 &\rightarrow &  A_{2u}+T_{1u}+T_{2u}\nonumber \\
J=4 &\rightarrow & A_{1u}+E_u+T_{1u}+T_{2u}.  \label{eq:cubicodd}
\end{eqnarray}
In Table~\ref{tab:Mmatrices-cubic-main} we have listed the cubic
symmetry labels of the spin matrices $\vec M_S$ and in
Table~\ref{tab:Nmatrices-cubic-main} those for the odd $S$ total
angular momentum matrices $\vec{N}_J$. [Correspondence to the
representations of the point group $T_d$ of the half-Heuslers in
provided in the Supp.\ Mat..]

An important property of discrete crystal point groups is that the
number of irreducible representations is finite. As a consequence,
distinct pairing channels labeled by different $J$ in full spherical
symmetry may contain several copies of the same cubic representation,
which implies that mixing is possible. 
This
is exemplified by Eq.~\eqref{eq:cubicodd}, from which we see that
e.g.\ certain $J=1,3,4$ pairing matrices can mix with one another
since all contain a representation with $T_{1u}$ symmetry. 

\section{Projection onto the valence bands}
\label{sec:proj-onto-valence}

As a preparatory step towards the derivation of the Eliashberg equation we now describe the process of projection to the states close to the Fermi energy. Since the electronic states relevant for the pairing instability are these states, it is natural to ignore pair scattering contributions which involve excitations at a higher energy scale, away from the Fermi surface. Usually, this is a trivial step where completely empty or completely filled bands are ignored without any consequence. However, in the present case, where spin-orbit coupling is so strong such that it splits the four fold multiplet in a way that one pair of bands folds upwards and the other downwards the projection will have an important effect.

The chemical potential, in this case, either crosses the hole-like
valence band ($\nu=-1$) or the electron-like conduction band ($\nu=+1$). 
In the case of hole-doping applicable to YPtBi, we then project out
the conduction band degrees of freedom and retain only the valence
band pair scattering terms of the interaction $V$. To this end, we
transform to the band basis and define the two-component valence band
electron operators $c_\bk$, which annihilate electrons in the
eigenstates $ |{\bf k} , \nu=-,\s \rangle$. The operators $c_\bk$ are related to the electron operators $\psi_\bk$ by
\be
c_{\mathbf{k}}=U^\dagger_{\bk} \psi_{\mathbf{k}},   \label{eq:34}
\ee
where $U_{\bk}$ is the $4\times 2$ matrix of valence band eigenvectors
(note that $c_\mathbf{k}$ and $U_\mathbf{k}$ in principle should carry
a $\nu$ index, but it is
left everywhere implicit, to avoid clutter). The projection operator $\mathcal{P}_{\nu}(\bk)$ onto the Kramers pair of bands denoted by $\nu$ takes the from
\be
\mathcal{P}_{\nu}(\bk) = \sum_{\s} |{\bf k} ,\nu,\s \rangle\langle {\bf k} ,\nu,\s| = U_{\bk}U^\dagger_{\bk}. \label{eq:projector1}
\ee
Projecting the irreducible pairing matrices $\vec M$ and $\vec N(\bk)$ onto the valence band basis yields $2\times 2$ pairing matrices, which we denote $\vec m(\bk)$ and $\vec n(\bk)$.
The latter are obtained from the $\vec M$ and $\vec N(\bk)$ matrices by
\be
 \vec m(\mathbf{k})=  U^\dagger_{\bk} \vec M U_{\bk}, \quad  \vec n(\mathbf{k})=  U^\dagger_{\bk} \vec N(\bk) U_{\bk}.  \label{eq:projectedpairing}
\ee
Note that generally lower case symbols denote the projected version of
the higher case ones (with their $\nu$ dependence suppressed).

The projection procedure performed by Eq.~\eqref{eq:projector1} can
also be expressed in a form which does not require choosing a basis
for the doubly degenerate valence band states. Using the Hamiltonian
of Eq.~\eqref{eq:36b} it is straightforward to establish that the
$4\times4$ form of the projection operator $\mathcal{P}_{\nu }(\bk)$ onto the $\nu$ bands is given by
\begin{equation}
{\rm P}_{\nu}(\bk) =\frac{1}{2}+\frac{\nu}{2E_{\bf k}}\left(c_1 \sum_{a=1}^3d_a(\mathbf{k})\Gamma_a+c_2 \sum_{a=4}^5d_a(\mathbf{k})\Gamma_a\right).    \label{eq:projector2}
\end{equation}
Note that in the presence of spherical symmetry (i.e.,
$c_1=c_2$ and $c_3=0$),
Eq.~\eqref{eq:projector2} simply becomes $\frac{1}{2}+\frac{\nu
  c_1}{2E_\mathbf{k}}\sum_{a=1}^5d_a(\mathbf{k})\Gamma_a$, and the
band {\em index} $\nu$ can be traded for $3/2$ ($\nu\,{\rm sign}c_1=1$) or
$1/2$ ($\nu\,{\rm sign}c_1=-1$), i.e.\ ${\rm P}_{3/2}(\bk) =\frac{1}{2}+\frac{
  |c_1|}{2E_\mathbf{k}}\sum_{a=1}^5d_a(\mathbf{k})\Gamma_a$ and ${\rm P}_{1/2}(\bk) =\frac{1}{2}-\frac{
  |c_1|}{2E_\mathbf{k}}\sum_{a=1}^5d_a(\mathbf{k})\Gamma_a$.


It is worth highlighting that the projection operators have the full
symmetry of the normal state system. 
Consequently, the representation labels---quantum number $J$ in
spherical symmetry---which characterize the irreducible pairing
channels remain good quantum numbers after projection. A remark
concerning the spin quantum number $S$ is in order, however. Within
the valence band, which is twofold pseudospin degenerate, only
pseudospin-singlet and pseudospin-triplet pairings can be formed. As a
result, Fermi statistics mandates that the $S=2$ and $S=3$ spin
pairing channels project onto the pseudospin-singlet
($\propto\sigma^0$) and pseudospin-triplet ($\propto\sigma^\mu$) channels, respectively. The multicomponent structure of the $S=2$ and $S=3$ spin pairing channels is then reflected in (additional) momentum dependence after projection onto the valence band. To see this in practice, consider the $(L,S)=(0,2)$ pairing channel. The five pairing matrices $\vec M_{J=S=2}$ simply project onto the five $d$-wave spherical harmonics $Y^{m}_{l=2}(\mathbf{\hat{k}})$, where $l$ is the orbital angular momentum. Specifically, the projected pairing matrices $\vec m_{S=2}(\mathbf{k})$ are given by
\be
 m^{m}_{S=2}(\mathbf{\hat{k}})= \pm Y^{m}_{l=2}(\mathbf{\hat{k}}) I_2.
\ee
In cubic symmetry, where $J=S=2$ splits into $E_g$ and $T_{2g}$, these projected pairings become
\be
m_{1,2,3}(\mathbf{\hat{k}})= \pm  c_1 \frac{d_{1,2,3}(\mathbf{k})}{ E_\bk}I_2 , \quad m_{4,5}(\mathbf{\hat{k}})= \pm c_2  \frac{d_{4,5}(\mathbf{k})}{ E_\bk}I_2.
\ee

We observe that, as a consequence of projecting onto the Fermi surface bands, only the parity of $S$ is a good quantum number. As a result, channels with equal $J$ but different $(L,S)$ can mix after projection. More specifically, if $\vec n_J (\bk)$ and $ \vec n'_J (\bk)$ are two sets of projected pairing matrices, obtained from channels with different $(L,S)$, they are not necessarily orthogonal. This mixing of channels with different spin and orbital quantum numbers can occur since projection onto the Fermi surface implies ignoring all pair scattering terms which involve the conductions band states. All inter-band and intra-conduction band pair scattering terms are projected out, and therefore, the information is retained is not sufficient to distinguish the quantum numbers $L$ and $S$.


In particular, this happens when projecting the channel with
non-trivial orbital angular momentum $L = 1$ and spin angular momentum
$S = 1$ and $S=3$. Both can form a total angular momentum $J = 2$. In
such cases we will explicitly add a label to the different {\it unprojected} representations, which project into the same representation $J$ by an additional index $j$, for example $N_{J=2,j}(\bf k)$ labels the two $S = 1$ and $S = 3$, which project to the same representation.

Now, inserting ${\rm Id}= \sum_\nu \mc P _\nu$ in Eqs.~(\ref{eq:ham}, \ref{eq:39}), and keeping only the
terms within a set of bands, and using the spherical symmetry
formulation, we obtain the effective action for the two bands which intercept the
Fermi energy:
\begin{multline}
\mathcal{S}^{\rm eff}=\sum_{k}c_{k}^\dagger
  (\mathsf{E}_\nu(\mathbf{k})-i\w)c_{k}\\
  +\frac{1}{8}\frac{1}{\beta \mc V}\sum_{\mathbf{k},\mathbf{k}',\w,\w'}\sum_J\hat{V}^{J}_{\alpha\beta\gamma\delta}(\mathbf{k},\mathbf{k}';\w-\w') \\
  \times c_{k \alpha }^\dagger c_{-k \beta}^\dagger c_{-k' \gamma}c_{ k' \delta}  \label{eq:36}
\end{multline}
where
\begin{eqnarray}
  \label{eq:37}
  &&\hat{V}^{J}_{\alpha\beta\gamma\delta}(\mathbf{k},\mathbf{k}';\w-\w')=\\
&&V_{J}(\w-\w')[\vec{n}_{J}(\mathbf{k})(i\sigma^y)]_{\alpha\beta}\cdot[(-i\sigma^y)\vec{n}_{J}^\dagger(\mathbf{k}')]_{\gamma\delta},\nonumber
\end{eqnarray}
where $V_{J}=V_{0,1}$ (from Eq.~\eqref{eq:38}) for $J$ coming from $S$
even or $S$ odd, respectively. Like in Eq.~\eqref{eq:40}, the sum over
$J$ runs over even and odd pairing channels, and
$\vec{n}_J(\mathbf{k})=\vec{m}_S$ for $S$ even. Eq.~\eqref{eq:37} is essentially Eq.~\eqref{eq:40} with the replacements
$M\rightarrow m$, $N\rightarrow n$, $\gamma\rightarrow(i\sigma^y)$,
$\psi\rightarrow c$.


As mentioned above, the sum over even $S$ matrices involves only the $2\times2$ identity matrix and can be written explicitly:
\be
\frac{1}{2}\left(1+\frac{c_1^2 \sum_ad_a(\mathbf{k})d_a(\mathbf{k}')}{E_\bk E_{\bk'}}\right)[c_{k}^\dagger i\sigma^y (c_{-k}^\dagger)^T][(i\sigma^yc_{-k'})^Tc_{k'}]
\ee
This (basis-dependent since the bands are degenerate) expressions is
useful to gain insight into the effect of the projection operators,
but in practice the actual diagonalization is not necessary, since
only the trace of the projected matrices appears in our calculations and we have the relation:
\begin{equation}
  \label{eq:50}
  {\rm Tr}[n^\alpha_{J}(\hat{\mathbf{k}}) n_{J'}^{\alpha'}{}^\dagger(\hat{\mathbf{k}})]={\rm
    Tr}[{\rm P}_\nu(\mathbf{k}) N^\alpha_{J}(\hat{\mathbf{k}}){\rm P}_\nu(\mathbf{k}) N_{J'}^{\alpha'}{}^\dagger(\hat{\mathbf{k}})].
\end{equation}
Therefore we will only formally assuming a diagonalization of the
Hamiltonian, but directly computing the right-hand-side of
Eq.~\eqref{eq:50} using the
general explicit expression Eq.~\eqref{eq:projector2}, which allows to perform
all analytical calculations.

\section{Linearized Eliashberg theory}
\label{sec:migd-eliashb-theory}

We are now in a position to analyze the superconducting instabilities
based on a general formalism for the derivation of the transition
temperature in spin-orbit coupled multiband systems with nontrivial structure. 
Our approach relies on Eliashberg
theory, the equations of which we derive from the lowest-order
self-energy correction due to the interaction, in the presence of superconducting test
vertices. Such a scheme corresponds to neglecting vertex corrections
at all orders and is equivalent to Dyson's equation truncated at first
order in the interaction.

Here, we will present the main steps of our analysis, relegating most
of the details to the Appendices. Furthermore, in our {\em presentation}, we
will consider spherical symmetry, and, for concreteness, focus specifically on a hole Fermi surface
(with pseudospin $\pm \frac32$ states), which is relevant for existing
experiments on YPtBi. 

\begin{figure}[htbp]
  \centering
  \includegraphics[width=\linewidth]{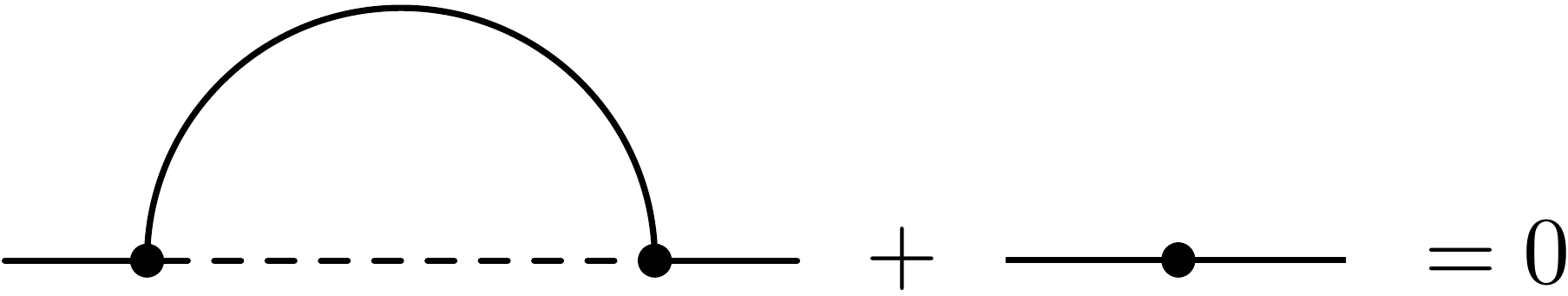}
  \caption{Diagrammatic representation of the Eliashberg equation. The solid lines are fermion propagators $\langle C_{k,a}C_{k,b}^\dagger\rangle_{\mathcal{S}_0'}$, where $C_{k}$ is the Nambu spinor of Eq.~\eqref{eq:52}. The dashed line represents the interaction $V$.
  \label{fig:eliashberg}}
\end{figure}

To obtain the Eliashberg equations starting from the projected effective action of Eq.~\eqref{eq:36}, we introduce a superconducting test vertex $\Sigma_A$. Specifically, we rearrange the normal part and interaction part of the action, $\mathcal{S}_0$ and $\mathcal{S}_{\rm int}$, as \cite{schrieffer1963theory}
\begin{eqnarray}
  \label{eq:43}
  \mathcal{S}_0&\rightarrow&\mathcal{S}_0'=\mathcal{S}_0-\mathcal{S}_A,\nonumber\\
\mathcal{S}_{\rm int}&\rightarrow&\mathcal{S}_{\rm int}'=\mathcal{S}_{\rm int}+\mathcal{S}_A,\nonumber
\end{eqnarray}
where the anomalous part $\mathcal{S}_A$ contains the test vertex $\Sigma_A = \Sigma_A(\bk,\w)$ and takes the form
\begin{equation}
  \label{eq:57}
  \mathcal{S}_A=\frac{1}{2}\sum_{k}\sum_{a,b} c_{k a}^\dagger (\Sigma_A i\sigma^y)_{ab}c_{-k b}^\dagger+{\rm h.c.}.
\end{equation}
Here, $a,b$ label the pseudospin degree of freedom $\pm \frac32$. (Recall that $k=(\bk,\w)$.) A self-consistent equation for the pairing test vertex $\Sigma_A$ is then obtained by setting $\langle \mathcal{S}_{\rm int}' \rangle_{\mathcal{S}_0'} = 0$, where $\langle X \rangle_{\mathcal{S}_0'}  \equiv \int DcDc^\dagger X e^{- \mathcal{S}_0'}$. Diagrammatically, the self-consistent equation $\langle \mathcal{S}_{\rm int}' \rangle_{\mathcal{S}_0'} = 0$ can be represented as in Fig.~\ref{fig:eliashberg}. Solving the self-consistent equation is then equivalent to solving a linearized gap equation for $T_c$.

For practical purposes it is convenient to adopt the Nambu spinor formalism and define
\be
C_{k}=\begin{pmatrix} c_{k}\\ i\sigma^y(c^\dagger_{-k})^T \end{pmatrix}.  \label{eq:52}
\ee
The normal part of the action $\mathcal{S}_0$ can then be expressed as
\begin{equation}
  \label{eq:53}
  \mathcal{S}_0= \frac{1}{2}\sum_{k}C_{k}^\dagger \begin{pmatrix}
\mathsf{E}_- -i\w & \\
 & -\mathsf{E}_- -i\w \end{pmatrix}C_{k},
\end{equation}
where $\mathsf{E}_-=\mathsf{E}_-(\mathbf{k})$ is the negative energy branch of the spectrum. The interaction part of the action takes the form
\begin{eqnarray}
  \label{eq:54}
&&  \mathcal{S}_{\rm int}=\frac{1}{32}\frac{1}{\beta
    \mc V}\sum_{k,k'}\sum_{J}V_{J}(\w-\w')\\
&&\quad\qquad\times(C_{k}^\dagger \vec{n}_{J}(\mathbf{k})\tau^+ C_{k})\cdot(C_{k'}^\dagger \vec{n}_{J}^\dagger(\mathbf{k}') \tau^-C_{k'}),\nonumber
\end{eqnarray}
where $\tau^\pm=\tau^x\pm i\tau^y$, and the Pauli matrices $\tau^{x,y,z}$ act on the Nambu spinor index. Finally, the anomalous part $\mathcal{S}_A$ takes the simple off-diagonal form
\begin{equation}
  \label{eq:56}
  \mathcal{S}_A=\frac{1}{2}\sum_{k}C_{k}^\dagger \begin{pmatrix} & \Sigma_A \\
\Sigma^\dagger_A  & \end{pmatrix}C_{k}.
\end{equation}
We can now calculate $\langle \mathcal{S}_{\rm int}'
\rangle_{\mathcal{S}_0'} $ using Wick's theorem, expressing the
quartic interaction in products of Nambu propagators $G^{ab}=  \langle
C_{k,a} C_{k,b}^\dagger \rangle_{\mathcal{S}_0'} $. The Green's
function $G$ has a matrix structure both in Nambu and pseudospin space, i.e.,
\be
  \label{eq:55}
 \begin{pmatrix} G_{11} &  G_{12} \\  G_{21}  & G_{22}   \end{pmatrix}=
\mathcal{K} \begin{pmatrix}\langle c_{k}c_{k}^\dagger \rangle & \langle c_{k} (c_{-k})^T \rangle \\
 \langle ( c_{-k}^\dagger)^T c_{k}^\dagger \rangle  & - \langle c_{-k}c_{-k}^\dagger\rangle^T  \end{pmatrix} \mathcal{K}^\dagger,
\ee
where $\mathcal{K}= \text{Diag}(1,i\sigma^y)$ and $\langle..\rangle=\langle..\rangle_{\mathcal{S}_0'}$. [Note that in
Eq.~\eqref{eq:55}, the matrix elements on the right-hand-side should
themselves be understood to be matrices: $\langle
c_{k}c_{k}^\dagger \rangle$ is for example to be read as $\langle
c_{k,a}c_{k,b}^\dagger \rangle$, and not as $\sum_a \langle
c_{k,a}c_{k,a}^\dagger \rangle$.] Since $\mathcal{S}_0'$ is quadratic
in the Nambu operators, the Green's function $G$ can be
straightforwardly found to be
\begin{equation}
  \label{eq:8}
G(\mathbf{k},\w)=\frac{i\w\tau^0+\mathsf{E}_- \tau^z-\Sigma_A \tau^x}{\w^2+\mathsf{E}_-^2+{\rm Tr}\Sigma_A^2},
\end{equation}
with the off-diagonal part given by
\begin{equation}
  \label{eq:48}
  G_{21}(\mathbf{k},\w)=-\frac{\Sigma_A  }{\w^2+\mathsf{E}_-^2+{\rm Tr}\Sigma_A^2}.
\end{equation}
Then, the linearized Eliashberg equation shown diagrammatically in Fig.~\ref{fig:eliashberg} takes the form
\begin{multline}
 \Sigma_A(\mathbf{k},\w)=
\frac{1}{4\beta \mc V}\sum_{\mathbf{k}',\w'}\sum_{J}\frac{V_{J}(\w-\w')}{\omega'{}^2+\mathsf{E}_-(\mathbf{k}')} \\
\times {\rm Tr} \left[G_{12}(\mathbf{k}',\w') (-i\sigma^y){n}^\dagger_{J}(\mathbf{k}')\right] {n}_{J}(\mathbf{k}).   \label{eq:6}
\end{multline}
Here we assumed a purely real pairing and assumed proximity to the
transition temperature where ${\rm Tr}\Sigma_A^2$ is small and could
be neglected. 

\subsection{Solving for $T_c$: Spherical symmetry}

\begin{table}[htbp]
  \centering
  \begin{tabular}{c|c|c|cc}
\hline\hline
  $\quad L\quad$  & $\quad S\quad$ & $\quad J\quad$ & $\qquad\left|{\bf k} ,{3/2},\s\right\rangle\qquad$ & $\qquad\left|{\bf k} ,{1/2},\s\right\rangle\qquad$ \\
\hline
0 &0 & 0 & $1/2$ & $1/2$ \\
0 &2 & 2 & $1/10$ & $1/10$ \\
\hline
1 &1 & 0 & ${\bf 9/10}$ & $1/10$ \\
1 &1 & 1 & $0$ & $2/5$ \\
1 &1,3 & 2 & $\tfrac{9}{25}\begin{pmatrix}1&\tfrac{-1}{\sqrt{14}}\\\frac{-1}{\sqrt{14}}&1/14\end{pmatrix}$ &
                                           $\bf {\tfrac{1}{25}\begin{pmatrix}\bf 7&\bf \tfrac{33}{\sqrt{14}}\\ \bf \tfrac{33}{\sqrt{14}}&{\bf  {177/14}}\end{pmatrix}}$\\
1 &3 & 3 & ${\bf 9/14}$ & $3/70$ \\
1&3 & 4 & $13/70$ & $27/70$ \\
\hline\hline
  \end{tabular}
  \caption{Strength of the projected pairing channels up to one power
    of $k$ in spherical symmetry [$O(3)$], $A_{J}=\frac{1}{4}\int
    {d\mathbf{\hat{k}}\over 4\pi}{\rm Tr}[{\rm P}_\nu(\mathbf{k}) N^\alpha_{J}(\mathbf{\hat{k}})
     {\rm P}_\nu(\mathbf{k}) N_{J}^\alpha{}^\dagger(\mathbf{\hat{k}})]$. $(L,S,J)$ stand for momentum (the power of $\bf k$), spin and their sum (i.e. total-angular momentum), respectively. Since we consider only local and single power of $\bf k$ pairing, only $L = 0$ and $L = 1$ appear (in principle $L$ can take all integer values).
    The parity (Even/Odd) of each channel is given by $(-1)^L$. 
    The bolded numbers mark the
    channels with highest non-s-wave pairing.
    Note that, after projection, the channels $(1,1,2)$ and $(1,3,2)$ mix. The corresponding
    matrix elements $A_{J}^{ii'}=\frac{1}{4}\int
    {d\mathbf{\hat{k}}\over 4\pi}{\rm Tr}[{\rm P}_\nu(\mathbf{k}) N^\alpha_{J,i}(\mathbf{\hat{k}})
    {\rm P}_\nu(\mathbf{k}) N_{J,i'}^{\alpha}{}^\dagger(\mathbf{\hat{k}})]$ are
    given on the fifth row of the table. The corresponding coupling
    strength is obtained by the largest eigenvalue of the matrix (see text).
    }
  \label{tab:strengths}
\end{table}

\begin{figure}[htbp]
  \centering
  \includegraphics[width=1.1\linewidth]{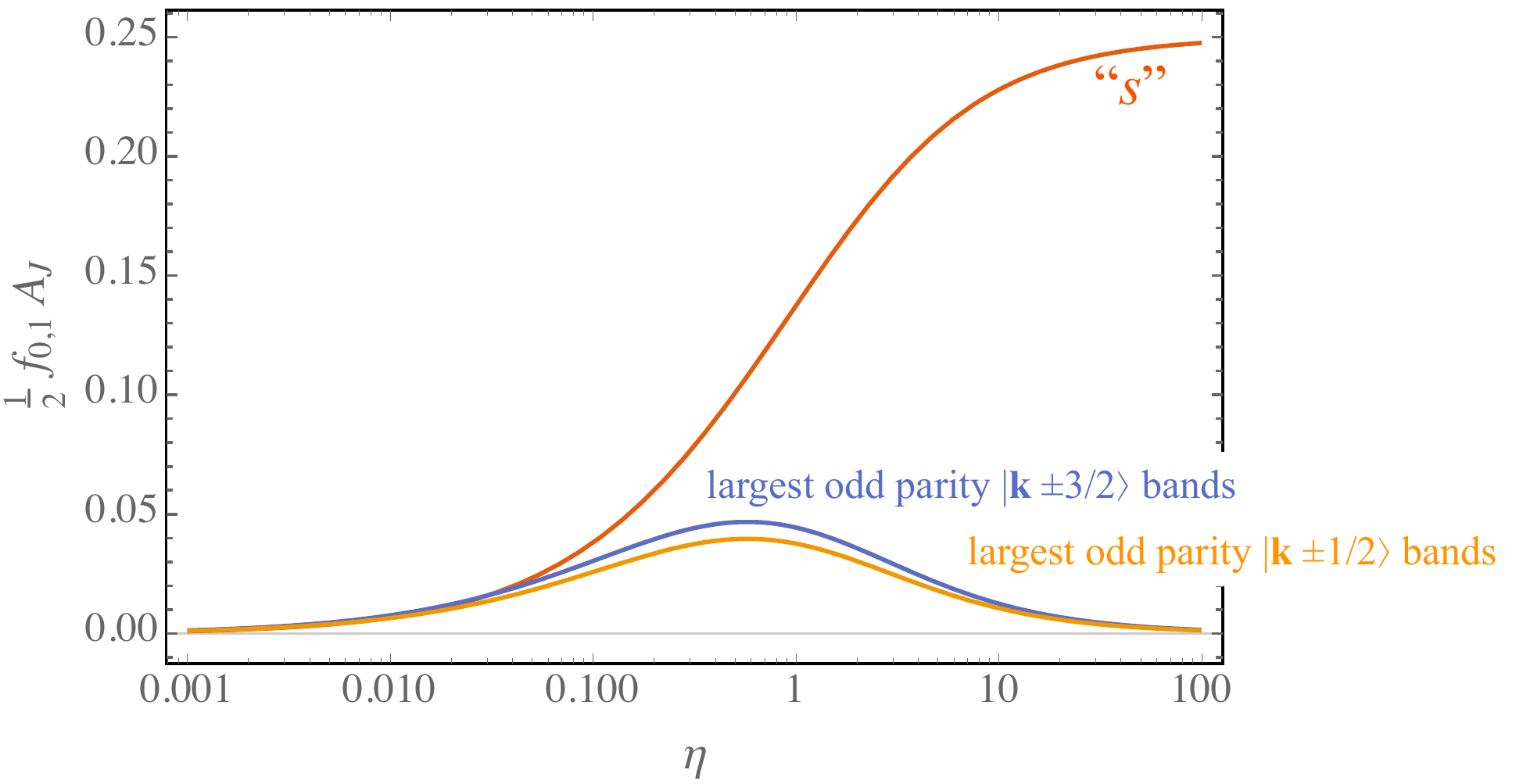}
  \caption{The value $\frac{1}{2}f_{0,1} A$ which controls the
    coupling strength of the Eliashberg equations as a function of the
    parameter $\eta$ defined in Eq.~(\ref{eta}) for the screened
    Coulomb interaction, $V(\mathbf{q},\w)={4 \pi e^2 /
      \left[\left(\ve_{\infty} + \ve_c(\w) \right)q^2 + 8\pi e^2 N(0)
      \right]}$. We assume spherical symmetry and particle-hole
    symmetric bands. For the usual {\em non}-spin-orbit coupled parabolic
    bands, the large-$\eta$ limit of $\frac{1}{2}f_0 A_0$ 
    is 0.5. The reduction to 0.25 for the
    spin-orbit coupled ``$s$'' pairing channel is responsible for the breakdown of Anderson's Theorem.}
\label{fig:strengths}
\end{figure}

Let us consider first the case of full spherical symmetry. We linearize the dispersion near the Fermi energy and perform the integration over momentum analytically.
As explained in Section \ref{sec:proj-onto-valence} there are in general two cases to consider. Let us first consider the simpler case, where the total angular momentum representation, $J$, derives from a unique set of quantum numbers $L$ and $S$. In that case, we may consider a pairing function of the form
$\Sigma_A(\mathbf{k},\w)=\Delta(\w)n^\alpha_J(\bf k)$
where $\Delta(\w)$ is a scalar and $\alpha=1,..,2J+1$. The Eliashberg equation then assumes the form
\begin{align}
  \label{eq:44}
\Delta_{J}(\w) =  -\frac{\pi}{\beta_{c,J}}\sum_{\w'}K_{J}(\w,\w')\Delta_{J}(\w')
\end{align}
where
\begin{align}
  \label{eq:45}
 &K_{J}(\w,\w')\\
&=\int \frac{d\mathbf{k}'}{(2\pi)^3}
    \frac{1}{4}V_{J}(\w-\w')\frac{{\rm Tr}[{\rm P}_-(\mathbf{k}') N^\alpha_{J}(\mathbf{k}')
    {\rm P}_-(\mathbf{k}')
    {N}^\alpha_{J}{}^\dagger(\mathbf{k}')]}{\w'^2+\mathsf{E}^2_-(\mathbf{k}')}\nonumber \\
    &= {A_{J}\over 2 |\w'|}{f_{0,1}\left[\eta(\w-\w')\right]} ,\nn
\end{align}
where the interaction $V$ was taken as in Eq.~\eqref{Coulomb} with the
electronic polarization function in the Thomas-Fermi approximation,
and the dispersion was linearized close to the Fermi surface. 

The strength of the attraction, encoded in the kernel $K_J$, is dictated by two factors. The first is the representation dependent constant:
\be
A_{J}=\frac{1}{4}\int{ d\mathbf{\hat{k}}\over 4\pi}{\rm Tr}[{\rm
  P}_\nu(\mathbf{k}) N^\alpha_{J}(\mathbf{\hat{k}}) {\rm
  P}_\nu(\mathbf{k}) N_{J}^\alpha{}^\dagger(\mathbf{\hat{k}})]\label{A_R}
  \ee
which can be found in Table~\ref{tab:strengths} for both $3/2$ and
$1/2$ bands (note that this constant is the same for all $\alpha=1,..,2J+1$ at
any given $J$). The second are the functions
\begin{eqnarray}
  \label{eq:29}
  f_0(\eta)&=&\frac{\eta}{2}\ln\left(1+\frac{2}{\eta}\right),\\
f_1(\eta)&=&\frac{\eta}{2}\left(-2+(2+\eta)\ln\left(1+\frac{2}{\eta}\right)\right).
\end{eqnarray}
Here $0$ or $1$ correspond to the even and odd representations of $J$, respectively, and the parameter $\eta$ quantifies the strength of the Coulomb interaction and is given by
\be
\eta(\w)={{q^2_{\rm TF}\left(\w\right)}\over{2k_{\rm F}^2}},
\label{eta}
\ee
 where $q_{\rm TF}(\w) = \sqrt{8\pi e^2 N(0) / [\ve_\infty + \varepsilon_c(\w)]}$ is the frequency-dependent Thomas-Fermi momentum. These two factors, $A_J$ and $f_{0,1}(\eta)$, multiplied together are plotted as a function of $\eta$ in Fig.~\ref{fig:strengths} for the three examples which give the highest coupling strength: $J = 0$ even (s-wave) in red, $J = 0$ odd in the hole band (p-wave) in blue, $J = 2$ odd in the electron band (p-wave) in orange.

From Eq.~(\ref{eq:45}) we find that the functions $f_{0,1}(\eta)$ depend on frequency only via the parameter $\eta$. At high frequencies the functions saturate to the value $f_{0,1}(\eta_\infty)$ and continuously go to $f_{0,1}(\eta_0)$ at zero frequency, where $\eta_{\infty} = \lim_{\w \rightarrow \infty}{{q^2_{\rm TF}(\w)}/{(2k_{\rm F}^2 )}}$ and $\eta_{0} = {{q^2_{\rm TF}(0)}/{(2k_{\rm F}^2 )}}$.

This allows us to understand how attraction appears, leading to
superconductivity, and to put bounds on
the transition temperature $T_c$ as follows. As is usual, the
interaction $V(\omega)$ can be decomposed into a static repulsion
$\mu$ (the high-frequency limit of the interaction) and an attractive
part $\lambda$ such that
\begin{equation}
  \label{eq:46}
  V_J(\omega)=\mu_J-\lambda_J(\omega),
\end{equation}
and
\begin{equation}
  \label{eq:47}
  \mu_J=\lim_{\omega\rightarrow+\infty}V_J(\omega),
\end{equation}
which also defines $\lambda_J(\omega)$.
Upon considering a low-energy theory, and therefore integrating out
large frequencies, the static repulsion $\mu_{J}$ 
is renormalized to a small dimensionless
repulsion $\mu_{J}^*$ so long as the Fermi energy is large compared
to the phonon frequency, which is the case for YPtBi. On the other
hand, $\lambda_J(\omega)$, which represents only the attraction due to
the electron-phonon interaction can be considered to be largely
unaffected by large-frequency effects. From here on
we set $\mu_J^*$ to be zero (as in standard BCS theory), so that now $V_J(\omega)\rightarrow
-\lambda_J(\omega)$. We can then read off the low frequency attractive
part of the interaction 
\begin{equation}
  \label{eq:49}
  \lambda_{J}=\lambda_J(\omega=0) =
\frac{A_{J}}{2}\left[f_{0,1}(\eta_\infty) -
  f_{0,1}(\eta_0)\right].
\end{equation}
Given
$\varepsilon_c(\infty)<\varepsilon_c(0)$, we have
$\eta_\infty>\eta_0$, so that
$\lambda_J>0$ in the $s$-wave channel as well as in the odd parity
channels if $\eta_\infty$ is not too large. The transition temperature in channel $J$ is then bound from above by $T_{c,J} < \w_L \exp\left[ -1/\lambda_{J} \right]$.

By estimating the coupling strengths in each symmetry
channel we find the following results:

\noindent

\noindent
{\em (i)} Looking at the first row of Table \ref{tab:strengths}, we
find that in the case of s-wave pairing the constant dictating the
coupling strength in the s-wave channel, $A_0$, is equal to
$1/2$. This should be compared with the analysis of
GLF \cite{gurevich1962}, where a simple quadratic band without
spin-orbit coupling was studied. In their case, calculating the same
constant gives $A_0 = 1$ (note that this would be true even if the
number of bands were to be multiplied by 2 to match the current
case). Thus, we find that the effectiveness of a local attraction in
generating s-wave pairing is dramatically reduced. This also implies
that Anderson's theorem does not hold in the case of a quadratic band
touching point. One way to see this is by considering the finely-tuned point $\a_2=\a_3=\a_4 = 0$ in
Eq.~(\ref{eq:ham}). There, all four bands
are degenerate, the Fermi energy therefore crosses all four bands, so that
no projection onto a subset of the latter should be performed, and $A_0
=1$. This simple example shows that the s-wave scattering matrix element can be modified by
tuning a parameter in the Hamiltonian without breaking time-reversal
symmetry: this is contrary to Anderson's conclusions in the case of
simple two-band systems (i.e.\ Kramers theorem still holds and therefore Anderson's does not).

\noindent
{\em (ii)} We find an explicit difference between the electron and hole bands in the odd-parity pairing coupling strengths as shown in
Table~\ref{tab:strengths}. The extreme example is the pairing in the
$J = 1$ representation, which is only allowed in the $1/2$ bands~\footnote{The physical reason for this is that the angular momentum difference between a pair of quasiparticles at $\bf k$ and $-\bf k$ in the same band $3/2$ and $\s$ is three units of angular momentum. Scattering between different momentum states of this type requires a higher angular momentum.}.

\noindent
{\em (iii)} The highest $T_c$ for non-trivial pairing in the $3/2$
bands (which is physically relevant for YPtBi) are the $L = 1$, $S =1$, $J = 0$ [corresponding to $N_0(\bf k) \propto \bf k \cdot \bf J$] and the $L = 1$, $S = 3$ and $J = 3$ [corresponding to Eq.~\eqref{eq:25}] channels, where the former is a one-dimensional representation and has a slightly higher coupling constant. In this case the maximal value the coupling constant can take is $\lambda \approx 0.05$. This is not enough to explain the transition temperature in YPtBi, for example, but is a non-negligible contribution. In the next section we discuss factors that may enhance the coupling in these two channels and favor them over the s-wave channel.

Finally, we note that in the case of $1/2$-band doping (electron doping in the YPtBi
case) the highest coupling constant is obtained in the case where two
sets of quantum numbers $L$ and $S$ mix, namely the case of of $L =
1$, $S = 1$ and $S = 3$, which both project into the odd $J = 2$ representation.
Let us now describe how to deal with this more complicated
situation. As explained in Section~\ref{sec:proj-onto-valence} the
group elements $N^\alpha_{2}(\bf k)$ are now labeled by an additional
index, $i$, which accounts for the $S = 1$ or $S = 3$ origin. The
constant Eq.~(\ref{A_R}) is then generalized to
 \be
A_{J}^{ii'}=\frac{1}{4}\int{ d\mathbf{\hat{k}}\over 4\pi}{\rm Tr}[{\rm
  P}_\nu(\mathbf{k}) N^\alpha_{J,i}(\mathbf{\hat{k}}) {\rm
  P}_\nu(\mathbf{k}) N_{J,i'}^\alpha{}^\dagger(\mathbf{\hat{k}})]\,\label{A_R,jj'}
  \ee
and forms a $2\times2$ matrix (see the fifth row in Table~\ref{tab:strengths}). $\Delta(\omega)n^\alpha_{J,i}(\mathbf{k})$ alone cannot solve the self-consistent equation Eq.~\eqref{eq:6}, but now a mixture of the two $n^\alpha_{2,i}(\mathbf{k})$ can be used. Defining
\begin{equation}
  \label{eq:3}
  \Sigma_A(k)=\Delta(\omega)\sum_{i=1,2}\phi_in^\alpha_{2,i}(\mathbf{k}),
\end{equation}
(i.e., not introducing additional $\mathbf{k}$ dependence in the
coefficients $\phi_i$), we find
that a set of solutions is given by solving for
the eigenvalues and eigenvectors of the matrix $\mathbf{A}_J=(A_j^{ii'})_{ii'}$ (the fifth row of
Table~\ref{tab:strengths})---see Supp.\ Mat.. For the $1/2$
bands, the largest coupling constant, i.e.\ that which will yield the
largest $T_c$, is
$\frac{1}{140}\left(55+\sqrt{2689}\right)\approx0.76$, and is obtained
for a pairing matrix equal to $0.59 n^\alpha_{J=2,S=1}+0.81
n^\alpha_{J=2,S=3}$. The other eigenvalues are given in
Table~\ref{tab:strengths}, and corresponding eigenvectors in the
Supp.\ Mat..




The {\em local} pairing states (with $L=0$ and $S=0,2$, i.e.\ rows
1,2,3 in Table~\ref{tab:Mmatrices-cubic-main}) were studied in detail
in Refs.~\cite{agterberg2017,timm2017,boettcher2017}. We leave the
analysis of the odd-parity pairing states from Table~\ref{tab:Nmatrices-cubic-main}), 
in particular the ones we find are favored by the polar-phonon mechanism, to future study.

\subsection{Factors that may favor non-s-wave pairing}

In the previous section we found that, in spite of a significant reduction, the density-density interaction Eq.~(\ref{Coulomb}) still favored s-wave pairing.
However, s-wave pairing is not consistent with recent penetration
depth measurements~\cite{kim2016}, which seem to indicate the
existence of nodes \cite{brydon2016,wang2017}.
In this short section we review corrections, which go beyond RPA, and may favor odd-parity pairing and enhance $T_c$.

First, we note that RPA relies on linear response. Namely, the response of the electronic polarization, taken into account in Eq.~\eqref{epsilon}, is taken to be linear. This breaks down at short distances much smaller than the screening core (of radius $r_{\rm TF} = 2\pi/q_{\rm TF}$), where the electric field becomes large. To correct for this, we consider an additional {\it local} interaction
\be
\d\mathcal{S}_{\text{int}}= \frac{\d V} 2 \int_{x}   \psi^\dagger_{x} \psi_{x} \psi^\dagger_{x} \psi_{x} ,   \label{eq:2}
\ee
When $\d V >0$ it enhances the repulsion, but only in the even parity
pairing channels (i.e. $L = 0$). Thus, in this case it favors p-wave
pairing due to enhanced local repulsion, i.e.\ by penalizing s-wave pairing.
However, even if the strength of the s-wave pairing is reduced, as we explained, this is not enough to account for the $T_c$ observed in experiment~\cite{butch2011superconductivity}, because the coupling in the $L = 1$, $S = 0$ and $J = 0$ channel is too weak.

The coupling strength can be enhanced when strong Fermi-liquid theory corrections are present. In particular the compressibility of a charged Fermi liquid is reduced by the Landau parameter $F_0^s$. As a result $q_{\rm TF}$ is also reduced and the interaction Eq.~(\ref{Coulomb}) is modified in the low frequency limit
\be
V(\w,\mathbf{q}) = { 1 + F_0 ^s\over 2 N(0)}{ \tilde q_{\rm TF}^2 (\w) \over q^2 +{\tilde q_{\rm TF}^2(\w)}}
\ee
where $\tilde q_{\rm TF}^2 (\w) =  q_{\rm TF}^2 (\w) / (1+ F_0
^s)$. Thus, the coupling strength is enhanced by a factor of $ 1 + F_0
^s$. Taking $\w_L  \approx 400$~K, we find that to explain the
measured $T_c = 0.77$~K in YPtBi one needs $F_0 ^s = 2.2$, which is a large, but not unrealistic, correction.

\subsection{Cubic symmetry}

Like before, our derivation carries over to cubic symmetry. In particular
Eq.~\eqref{eq:6} and the first equality of Eq.~\eqref{eq:45} are still
valid with the replacement of the index $J$ by the cubic
representations listed in Tables~\ref{tab:Mmatrices-cubic-main} and
\ref{tab:Nmatrices-cubic-main}. However, no simple form such as the
second equality of Eq.~\eqref{eq:45} exists in that case. Indeed, there, we made use of
the isotropy of $\mathsf{E}_-$---which is no longer true in cubic
symmetry. All of the angular dependence was then carried only by the factor
${\rm Tr}[{\rm P}_-(\mathbf{k}')N_J^\alpha(\mathbf{k}') {\rm
  P}_-(\mathbf{k}')N_J^\alpha{}^\dagger(\mathbf{k}')]$, which itself did
not depend on the magnitude of $\mathbf{k}'$, but only its direction. In cubic symmetry, no
such trivial separation of the dependences on the direction and
magnitude of $\mathbf{k}$ exists. In that case, the coefficients $A_R$
will carry no real meaning, and one needs to resort to
numerical estimates of the full $d\mathbf{k}'$ integral for each set
of parameter values. Physically, because of the additional angular
dependence in the integral, one expects this will typically tend to enhance the odd
parity pairings, but it seems not enough to overcome that of the
$s$-wave.

It is also worth noting that, in cubic symmetry,
mixing of several representation copies is the rule rather than the
exception (as was the case in spherical symmetry where only $J=2,L=1$
and $S=1,3$ mixed), because most representations appear several times.

\section{Discussion}
\label{sec:discussion}
We have presented a theory for the study of superconductivity in spin-orbit coupled materials and applied it to $j=3/2$ semimetals in three dimensions.
We used our theory to study the pairing strength due to a polar optical phonon as first discussed by GLF~\cite{gurevich1962}. We showed that the coupling strength can potentially be large enough to explain superconductivity in the half-Heuslers, in contrast to the conclusion of Ref.~\onlinecite{meinert2016}.

Furthermore, we have classified all possible pairing channels, which are local or linear in momentum. Within RPA we found that the highest $T_c$ was in the s-wave channel, and that there were multiple odd-parity channels with comparable pairings. As we pointed out, corrections which go beyond linear response may favor pairing in these channels and account for the $T_c$ observed in experiments. It is important to note however, that the full dynamical and momentum dependence of the dielectric constant Eq.~(\ref{epsilon}) needs to be taken into account to be able to make better estimates of the coupling constants. We leave this to future study.

We also point out that our study led us to a few more general
results. First we found that the coupling strength in the s-wave
channel was significantly reduced in the quadratic band touching case,
which is the direct result of projecting out the unoccupied electron
bands and invalidates Anderson's theorem for these kind of
semimetals. We also showed that the pairing strength and the resulting
expected gap symmetry was different in the case of electron and hole
doping. Thus, we expect that the superconducting state in an
electron-doped half-Heusler will be different than in the hole doped ones.


\section{Acknowledgements}

We thank Leon Balents, Max Metlitski and Doug Scalapino for useful
discussions. LS also acknowledges Leon Balents and Eun-Gook Moon for a
prior collaboration on a related topic. LS and JR were supported by the
Gordon and Betty Moore Foundation through scholarships of the EPiQS
initiative under grant no.\ GBMF4303. LS also acknowledges the
hospitality of the KITP, where part of this work was carried out, and
NSF grant PHY-1125915. PAL was supported by the DOE under grant no.\ FG02-03ER46076. LF and JWFV acknowledge funding by the DOE Office of
Basic Energy Sciences, Division of Materials Sciences and Engineering under Award No. de-sc0010526.

\bibliography{YPtBi-bib}

\appendix

\section{Definitions and parameter values}
\label{sec:definitions}

The Fourier and Matsubara transformations ($\omega$ is a fermionic
Matsubara frequency, $\omega=\omega_n=\frac{(2n+1)\pi}{\beta}$) are carried out using the following normalization
\begin{equation}
  \label{eq:27}
  \psi_x=\frac{1}{\sqrt{\beta}\sqrt{\mathcal{V}}}\sum_{\omega,\mathbf{k}}e^{i\mathbf{k}\cdot\mathbf{r}-i\omega\tau}\psi_{k}.
\end{equation}

\subsection{Hamiltonian definitions}
\label{sec:hamilt-defin}

The fermionic Hamiltonian density reads
\begin{eqnarray}
\mathcal{H}_0(\mathbf{k})&=&\alpha_1 \mathbf{k}^2 + \alpha_2
                             \left(\mathbf{k} \cdot
                             \mathbf{J}\right)^2 + \alpha_3
                             \left(k_x^2 J_x^2 + k_y^2 J_y^2 +k_z^2
                             J_z^2\right) \nonumber \\
&&+\alpha_4\mathbf{k}\cdot\mathbf{T}-\mu  \label{eq:lut}\\
&=&c_0 \mathbf{k}^2+ \sum_{a=1}^5  \hat{c}_a d_a(
    \mathbf{k})\Gamma_a+c_3\mathbf{k}\cdot\mathbf{T}-\mu,
\label{eq:eqcs}
\end{eqnarray}
where $\hat{c}_1=\hat{c}_2=\hat{c}_3=c_1$ and
$\hat{c}_4=\hat{c}_5=c_2$.
The first line uses the conventional Luttinger parameters
($\alpha_{1,2,3}$) in the $j=3/2$ matrix representation
\cite{luttinger1956}, and the
second line is the form used in the main text. The Gamma
matrices ($\Gamma_{a}$) form a Clifford algebra,
$\{\Gamma_a,\Gamma_b\}=2\delta_{ab}$, and have been introduced as described in the literature
\cite{murakami2004}, and
\begin{eqnarray}
&&d_1( \mathbf{k})=\frac{k_xk_y}{\sqrt{2}},\quad d_2( \mathbf{k})=\frac{k_xk_z}{\sqrt{2}},\quad d_3( \mathbf{k})=\frac{k_yk_z}{\sqrt{2}} \nonumber \\
&&d_4( \mathbf{k})=\frac{k_x^2-k_y^2}{2\sqrt{2}},\quad d_5( \mathbf{k})=\frac{2k_z^2 -k_x^2-k_y^2}{2\sqrt{6}}. \nonumber
\end{eqnarray}
Note that $c_0$ ($\alpha_1$) quantifies the particle-hole
asymmetry, while $\left|c_1-c_2\right|$ ($\alpha_3$) naturally characterizes the
cubic anisotropy and $c_3$ ($\alpha_4$) the departure from inversion symmetry. In the absence of inversion breaking, i.e.\ when $c_3=0$, the energy eigenvalues are $\mathsf{E}_{\pm}(\mathbf{k}) = c_0 \mathbf{k}^2 \pm
E(\mathbf{k})-\mu$, where $E(\mathbf{k})=\sqrt{\sum_{a=1}^5 \hat{c}_a^2  d_a
  ^2( \mathbf{k}) }$ and the Hamiltonian density can be rewritten
\begin{equation}
\mathcal{H}^{\rm
  inv}_0(\mathbf{k})=\sum_{\nu=\pm1}\mathsf{E}_\nu(\mathbf{k}){\rm P}_\nu(\mathbf{k}),
\end{equation}
 where ${\rm P}_{\nu}(\mathbf{k}) = \frac{1}{2}\left(1+\nu
\frac{\mathcal{H}_0(\mathbf{k})-c_0\mathbf{k}^2 +\mu}{E(\mathbf{k})}\right)$
is a projection operator, ${\rm P}_{\nu}^2(\mathbf{k})={\rm
  P}_{\nu}(\mathbf{k})$ (no summation).

It is straightforward to relate the $c_i$ coefficients used in Eq.~\eqref{eq:eqcs} to
the Luttinger $\alpha_i$ parameters used in Eq.~\eqref{eq:lut}. This can be done by expressing
the spin operators in terms of the Gamma matrices, using for example the equalities
\begin{eqnarray}
&& J_x = \frac{\sqrt{3}}{2} \Gamma_{15} - \frac{1}{2} (\Gamma_{23} - \Gamma_{14}) \ ,\nonumber \\
&& J_y = -\frac{\sqrt{3}}{2} \Gamma_{25} + \frac{1}{2} (\Gamma_{13} + \Gamma_{24}) \ ,\nonumber \\
&& J_z = - \Gamma_{34} - \frac{1}{2} \Gamma_{12} \ ,
\end{eqnarray}
where $\Gamma_{ab} = \frac{1}{2 i}
[\Gamma_a, \Gamma_b]$. We find
\begin{equation}
  \label{eq:63}
\begin{cases}
  c_0=\alpha_1+\frac{5}{4}(\alpha_2+\alpha_3)\\
c_1=\sqrt{6}\alpha_2\\
c_2=\sqrt{6}(\alpha_2+\alpha_3)\\
c_3=\alpha_4
\end{cases},\;\mbox{i.e.}\quad
\begin{cases}
\alpha_1=c_0-\frac{5}{4\sqrt{6}}c_2\\
\alpha_2=\frac{c_1}{\sqrt{6}}\\
 \alpha_3=\frac{c_2-c_1}{\sqrt{6}}\\
\alpha_4=c_3
\end{cases}.
\end{equation}

Note that if explicit matrices are used, they follow the definitions in
Ref.~\cite{murakami2004}. For these definitions, the $4\times4$ antisymmetric
matrix $\gamma$ used throughout is equal to $\gamma=-i\Gamma_{13}$.

Finally, the transformation of the $d_a$ under a three-fold rotation
around the $[111]$ axis is:
\begin{eqnarray}
&&d_1\rightarrow d_2\rightarrow d_3\rightarrow d_1,\\
&&d_4\rightarrow\frac{-1}{2}(d_4+\sqrt{3}d_5),\quad d_5\rightarrow\frac{1}{2}(\sqrt{3}d_4-d_5).\nonumber
\end{eqnarray}
$\Gamma_a$ transforms like $d_a$.

\subsubsection{Spherical symmetry}
\label{sec:spherical-symmetry}

In spherical symmetry, $c_1=c_2=c$ and $c_3=0$, and
$\alpha_3=\alpha_4=0$. Also the $|j^z=\pm1/2\rangle$ and
$|j^z=\pm3/2\rangle$ are good eigenstates, with eigenenergies
\begin{eqnarray}
  \label{eq:16}
  \mathsf{E}_{1/2}(\mathbf{k})&=&\mathbf{k}^2(\alpha_1+\frac{1}{4}\alpha_2)=\mathbf{k}^2(c_0-\frac{c}{\sqrt{6}}),\\
\mathsf{E}_{3/2}(\mathbf{k})&=&\mathbf{k}^2(\alpha_1+\frac{9}{4}\alpha_2)=\mathbf{k}^2(c_0+\frac{c}{\sqrt{6}}).
\end{eqnarray}
In terms of hole and electron bands,
\begin{equation}
  \label{eq:65}
  \mathsf{E}_\pm(\mathbf{k})=\left(c_0\pm\frac{|c|}{\sqrt{6}}\right)\mathbf{k}^2.
\end{equation}
From these equations, we find the relations between $3/2$ and $1/2$
bands and $\nu=\pm1$ electron and hole bands, in spherical symmetry:
\begin{equation}
  \label{eq:66}
\begin{cases}
\nu\,{\rm sign}c=-1\;\Leftrightarrow\;1/2\\
  \nu\,{\rm sign}c=+1\;\Leftrightarrow\;3/2
\end{cases}.
\end{equation}

\subsection{Parameter definitions}
\label{sec:param-defin}
In Gaussian units ($\pm$ refers to eletron/hole bands), $m$ the
effective mass, $k_{\rm F}$ the Fermi energy, $n$ the carrier density,
$N(0)$ the density of states at the Fermi energy, $a_0$ the effective
Bohr radius, $q_{\rm TF}$ the Thomas-Fermi momentum, ${\rm Ry}$ the
effective Rydberg, and $E_{\rm F}=|\mu|$ the Fermi energy are:
\begin{equation}
  \label{eq:13}
\begin{cases}
m=\frac{\hbar^2}{2(c_0\pm|c|/\sqrt{6})}\\
k_{\rm F}=\frac{\sqrt{2m
    E_F}}{\hbar}=\sqrt{\frac{E_F}{c_0\pm|c|/\sqrt{6}}}\\
n=\frac{1}{(2\pi)^3}\frac{4\pi}{3}k_{\rm F}^3=\frac{1}{6\pi^2}\sqrt{\frac{E_F}{c_0\pm|c|/\sqrt{6}}}^3\\
N(0)=\frac{mk_F}{2\pi^2\hbar^2}=\frac{k_F}{4\pi^2(c_0\pm|c|/\sqrt{6})}=\frac{\sqrt{E_F}}{4\pi^2(c_0\pm|c|/\sqrt{6})^{3/2}}\\
a_0=\frac{\hbar^2}{me^2}=\frac{2(c_0\pm|c|/\sqrt{6})}{e^2}\\
q_{\rm TF}(\w)=\sqrt{8\pi
  e^2N(0)/\ve_c(\w)}=\sqrt{\frac{e^2}{\pi \ve_c(\w)}\frac{\sqrt{E_F}}{(c_0\pm|c|/\sqrt{6})^{3/2}}}\\
{\rm Ry}=\frac{\hbar^2}{2m a_0^2}=\frac{me^4}{2\hbar^2}=\frac{e^4}{4(c_0\pm|c|/\sqrt{6})}
\end{cases}.
\end{equation}

\subsection{Parameter values}
\label{sec:parameter-values}

Yet another notation for the Hamiltonian density is used in Ref.~\onlinecite{brydon2016},
\begin{eqnarray}
\mathcal{H}_0(\mathbf{k})&=&\alpha\mathbf{k}^2+\beta(k_x^2J_x^2+k_y^2
J_y^2 +k_z^2 J_z^2)+\gamma\sum_{\mu\neq\nu}k_\mu k_\nu J_\mu
J_\nu\nonumber\\
&&+\delta\sum_\mu k_\mu(J_{\mu+1}J_\mu J_{\mu+1}-J_{\mu+2}J_\mu J_{\mu+2})-\mu,\nonumber\\
\end{eqnarray}
which yields 
\begin{equation}
  \label{eq:67}
\begin{cases}
  c_0=\alpha+\frac{5}{4}\beta\\
c_1=\sqrt{6}\gamma\\
c_2=\sqrt{6}\beta\\
 c_3=\frac{\sqrt{3}}{2}\delta
\end{cases},\;\mbox{i.e.}\quad
\begin{cases}
\alpha=c_0-\frac{5}{4\sqrt{6}}c_2\\
\beta=\frac{c_2}{\sqrt{6}}\\
\gamma=\frac{c_1}{\sqrt{6}}\\
\alpha_4=\frac{2c_3}{\sqrt{3}}
\end{cases}.
\end{equation}

Plugging in the values given for YPtBi in the caption of Fig.~2 of
Ref.~\onlinecite{brydon2016}, i.e.\ 
\begin{equation}
  \label{eq:68}
  \begin{cases}
\alpha=20.(a/\pi)^2\;{\rm eV}\\
\beta=-15.(a/\pi)^2\;{\rm eV}\\
\gamma=-10.(a/\pi)^2\;{\rm eV}\\
\delta=0.1(a/\pi)^2\;{\rm eV} 
\end{cases},\;\mbox{so}\;
\begin{cases}
c_0=1.25(a/\pi)^2\;{\rm eV}\\
c_1=-24.5(a/\pi)^2\;{\rm eV}\\ 
c_2=-36.7(a/\pi)^2\;{\rm eV}\\
c_3=0.0866(a/\pi)^2\;{\rm eV}
\end{cases},
\end{equation}
and $\mu=-20$~meV,
we obtain $|c_0/c_1|=0.051<1/\sqrt{6}$ indeed, as well as, taking for a spherical
approximation $c_3=0$ and $c_1=c_2=c\approx-30.6(a/\pi)^2$~eV, and the lattice
constant $a=6.65\,10^{-10}$~m,
\begin{equation}
  \label{eq:14}
  \begin{cases}
m=7.5\,10^{-2}m_e=6.83\,10^{-32}\,{\rm kg}\\
k_{\rm F}=2.0\,10^{8}\,{\rm m}^{-1}\\
n=1.33\,10^{23}\;{\rm m}^{-3}=1.33\,10^{17}\;{\rm cm}^{-3}\\
N(0)=1.00\,10^{25}\;{\rm eV}^{-1}{\rm m}^{-3}\\
a_0=13.3a_B=7.01\,10^{-10}\,{\rm m}\\
q_{\rm TF}=4.3\,10^{8}\,{\rm m}^{-1}\\
{\rm Ry}=7.5\,10^{-2}{\rm Ry}_0=1.03\,{\rm eV}\\
E_{\rm F}/{\rm Ry}=1.9\,10^{-2}\\
q_{\rm TF}/k_{\rm F}=2.1\\
\eta=\frac{q_{\rm TF}^2}{2k_{\rm F}^2}=2.3\\
N(0)a_0^3=3.4\,10^{-3}\;{\rm eV}^{-1}\\
N(0)/k_{\rm F}^3=1.3\;{\rm eV}^{-1}\\
N(0)/q_{\rm TF}^3=0.13\;{\rm eV}^{-1}
\end{cases},
\end{equation}
where $m_e$ is the electron mass, $a_B$ the Bohr radius, and Ry$_0$ the
Rydberg. Note that with these values (and in the spherical approximation taken with
$c=(c_1+c_2)/2$), we obtain a density $n=1.33\,10^{17}$~cm$^{-3}$, smaller
than the one reported experimentally, $n\sim10^{18}$~cm$^{-3}$.

\section{Matrices and pairings}
\label{sec:matrices-pairings}

In this section and associated tables, all matrices are orthonormalized according to the following scalar product
\begin{equation}
  \label{eq:18}
  (\mathcal{M}|\mathcal{N})=\frac{1}{4\pi}\int
  d\mathbf{\hat{k}}{\rm Tr}[\mathcal{M}(\mathbf{\hat{k}})\mathcal{N}^\dagger(\mathbf{\hat{k}})],
\end{equation}
where $\mathcal{M}$ and $\mathcal{N}$ are $4\times4$ matrices that may
or may not depend on $\mathbf{\hat{k}}$, and a matrix $\mathcal{M}$ is
normalized if $(\mathcal{M}|\mathcal{M})=4$.

For convenience, we define
${\boldsymbol{\mathcal{J}}}=\frac{-41}{6\sqrt{5}}\mathbf{J}+\frac{2\sqrt{5}}{3}(J_x^3,J_y^3,J_z^3)$. Note
that $(\mathcal{J}^\mu|J^\nu)=0$ $\forall \mu,\nu$.

\subsection{Spherical symmetry}
\label{sec:spherical-symmetry-1}

\begin{table}[htbp]
  \centering
  \begin{tabular}[c]{ccc}
\hline\hline
$S$ & $\vec{M}_S$ & par. \\
\hline
$0$ & $I_4$ & Even\\
$2$ & $\frac{1}{\sqrt{2}}(-i\Gamma_3-\Gamma_4,i\Gamma_1+\Gamma_2,-\sqrt{2}\Gamma_5,i\Gamma_1-\Gamma_2,i\Gamma_3-\Gamma_4)$ & Even \\
$1$ & $\frac{\sqrt{2}}{\sqrt{5}}(J_x+iJ_y,-\sqrt{2}J_z,-J_x+iJ_y)$ & Odd\\
$3$ & $(M_3^3,M_3^2,M_3^1,M_3^0,M_3^{-1},M_3^{-2},M_3^{-3})$: see below & Odd\\
\hline\hline
  \end{tabular}
  \caption{Matrices $M_S^\alpha$ in spherical symmetry. The 
    column ``par.'' indicates whether $S$ is even or odd (i.e.\
    whether $M_S\gamma$ is antisymmetric or
    symmetric, respectively).}
  \label{tab:}
\end{table}
\begin{equation}
  \label{eq:19}
  \begin{cases}
    M_3^{3}=\frac{1}{2}(-i\Gamma_{13}-\Gamma_{14}-\Gamma_{23}+i\Gamma_{24})\\
    M_3^{2}=\frac{1}{\sqrt{2}}(-\Gamma_{35}+i\Gamma_{45})\\
    M_3^{1}=\frac{\sqrt{3}}{2\sqrt{5}}(-i\Gamma_{13}-\Gamma_{14}+\frac{2}{\sqrt{3}}\Gamma_{15}+\Gamma_{23}-i\Gamma_{24}-\frac{2i}{\sqrt{3}}\Gamma_{25})\\
    M_3^{0}=\frac{1}{\sqrt{5}}(2\Gamma_{12}-\Gamma_{34})\\
    M_3^{-m}=(M_3^m)^{\dagger}\;\;\forall\; m
  \end{cases}
\end{equation}
$\mathbf{k}$ transforms as $L=1$ for $SO(3)$ operations.
\begin{table}[htbp]
  \centering
  \begin{tabular}[c]{ccc}
\hline\hline
$J$ & $S$& $\vec{N}_{J}(\mathbf{k})$ \\
\hline
$0$ &1& $\frac{2}{\sqrt{5}}\mathbf{k}\cdot\mathbf{J}$  \\
$1$ &1& $(N_{1}^1,N_{1}^0,N_{1}^{-1})$: see below \\
$2$ &1&
      $(N_{2^{(1)}}^2,N_{2^{(1)}}^1,N_{2^{(1)}}^0,N_{2^{(1)}}^{-1},N_{2^{(1)}}^{-2})$: see below \\
\hline
$2$ &3& $(N_{2^{(3)}}^2,N_{2^{(3)}}^1,N_{2^{(3)}}^0,N_{2^{(3)}}^{-1},N_{2^{(3)}}^{-2})$: see below \\
$3$ &3& $(N_3^3,N_3^2,N_{3}^1,N_{3}^0,N_{3}^{-1},N_3^{-2},N_3^{-3})$: see below \\
$4$ &3& $(N_4^4,N_4^3,N_4^2,N_{4}^1,N_{4}^0,N_{4}^{-1},N_4^{-2},N_4^{-3},N_4^{-4})$: see below \\
\hline\hline
  \end{tabular}
  \caption{Odd parity pairing matrices
    $N^\alpha_{J}(\mathbf{k})$ with a single power of $\mathbf{k}$ in spherical symmetry.}
  \label{tab:}
\end{table}
\begin{equation}
  \label{eq:22}
  \begin{cases}
    N_1^1(\mathbf{k})=\frac{\sqrt{3}}{\sqrt{5}} (-k_z(J_x+iJ_y)+(k_x+ik_y)J_z)\\
N_1^0(\mathbf{k})=i\frac{\sqrt{6}}{\sqrt{5}}(k_yJ_x-k_xJ_y)
  \end{cases}
\end{equation}
\begin{equation}
  \label{eq:23}
  \begin{cases}
N_{2^{(1)}}^2(\mathbf{k})=\frac{\sqrt{3}}{\sqrt{5}} ((k_x+ik_y)(J_x+iJ_y))\\
N_{2^{(1)}}^1(\mathbf{k})=\frac{\sqrt{3}}{\sqrt{5}} (-k_z(J_x+iJ_y)-(k_x+ik_y)J_z)\\
N_{2^{(1)}}^0(\mathbf{k})=i\frac{\sqrt{2}}{\sqrt{5}}(-k_xJ_x-k_yJ_y+2k_zJ_z)
  \end{cases}
\end{equation}
\begin{equation}
  \label{eq:24}
  \begin{cases}
    N_{2^{(3)}}^2(\mathbf{k})=\frac{\sqrt{3}}{\sqrt{5}} (-k_z(J_x+iJ_y)+(k_x+ik_y)J_z)\\
    N_{2^{(3)}}^1(\mathbf{k})=\frac{\sqrt{3}}{\sqrt{5}} (-k_z(J_x+iJ_y)+(k_x+ik_y)J_z)\\
N_{2^{(3)}}^0(\mathbf{k})=i\frac{\sqrt{6}}{\sqrt{5}}(k_yJ_x-k_xJ_y)
  \end{cases}
\end{equation}
\begin{equation}
  \label{eq:25}
  \begin{cases}
N_3^3(\mathbf{k})=
\frac{1}{2}Y_{11}M_3^2-\frac{\sqrt{3}}{2}Y_{10}M_3^3
\\
N_3^2(\mathbf{k})=\sqrt{\frac{5}{12}}Y_{11}M_3^1-\sqrt{\frac{1}{3}}Y_{10}M_3^2-\frac{1}{2}Y_{1-1}M_3^3\\
N_3^1(\mathbf{k})=\frac{1}{\sqrt{2}}Y_{11}M_3^0-\frac{1}{2\sqrt{3}}Y_{10}M_3^1-\sqrt{\frac{5}{12}}Y_{1-1}M_3^2\\
N_3^0(\mathbf{k})=\sqrt{\frac{1}{2}}Y_{11}M_3^{-1}-\sqrt{\frac{1}{2}}Y_{1-1}M_3^1\\
  \end{cases}
\end{equation}
\begin{equation}
  \label{eq:26}
  \begin{cases}
N_4^4(\mathbf{k})=Y_{11}M_3^3\\
N_4^3(\mathbf{k})=\frac{\sqrt{3}}{2}Y_{11}M_3^3+\frac{1}{2}Y_{10}M_3^3\\
N_4^2(\mathbf{k})=\sqrt{\frac{15}{28}}Y_{11}M_3^1-\sqrt{\frac{3}{7}}Y_{10}M_3^2+\frac{1}{2\sqrt{7}}Y_{1-1}M_3^3\\
N_4^1(\mathbf{k})=\sqrt{\frac{5}{14}}Y_{11}M_3^0+\sqrt{\frac{15}{28}}Y_{10}M_3^1+\sqrt{\frac{5}{28}}Y_{1-1}M_3^2\\
N_4^0(\mathbf{k})=\sqrt{\frac{3}{14}}Y_{11}M_3^{-1}+\sqrt{\frac{4}{7}}Y_{10}M_3^0+\sqrt{\frac{3}{14}}Y_{1-1}M_3^1\\
  \end{cases},
\end{equation}
where the $Y_{lm}(\mathbf{\hat{k}})$ are the usual spherical
harmonics, normalized following $\frac{1}{4\pi}\int d\mathbf{\hat{k}} Y_{lm}^*(\mathbf{\hat{k}})Y_{lm}(\mathbf{\hat{k}})=1$ (and we have
switched in Eqs.~(\ref{eq:25},\ref{eq:26}) from the $k^\mu$ to the
spherical harmonic notation for compactness).

\subsection{Cubic symmetry $O_h$}
\label{sec:cubic-symmetry}

In which ``form'' we write down the matrices ($\Gamma_a$ or $J^\mu$)
in the tables and equations is entirely determined by the simplest form.

\begin{table}[htbp]
  \centering
  \begin{tabular}[c]{cccc}
\hline\hline
$R$ & $\vec{M}_R$ & par. & $R(T_d)$ \\
\hline
$A_{1g}$ & $I_4$ & Even & $A_{1}$\\
$E_g$ & $(\Gamma_4,\Gamma_5)$ & Even & $E$\\
$T_{2g}$ & $(\Gamma_1,\Gamma_2,\Gamma_3)$ & Even & $T_{2}$\\
$T_{1g}$ & $\frac{2}{\sqrt{5}}(J_x,J_y,J_z)$ & Odd & $T_{1}$\\
$A_{2g}$ & $\frac{2}{\sqrt{3}}(J_xJ_yJ_z+J_zJ_yJ_x)=-\Gamma_{45}$ &
                                                                    Odd
                   & $A_{2}$\\
$T_{1g}$ &
           $\frac{-41}{6\sqrt{5}}\mathbf{J}+\frac{2\sqrt{5}}{3}(J_x^3,J_y^3,J_z^3)$
            & Odd & $T_{1}$ \\
$T_{2g}$ & $\frac{-1}{\sqrt{3}}(T_x,T_y,T_z)$ & Odd & $T_{2}$\\
\hline\hline
  \end{tabular}
  \caption{Matrices $\vec{M}_R$ in cubic symmetry with inversion $O_h$, and
    in tetrahedral symmetry $T_d$ (where one simply reads the
    representation labels with the $g$ index dropped). The parity
    column ``par.'' indicates whether $M_R\gamma$ is symmetric (Odd) or
    antisymmetric (Even).}
  \label{tab:Mmatrices-cubic}
\end{table}

$\mathbf{k}$ transforms under the $T_{1u}$ representation of $O_h$.
\begin{table}[htbp]
  \centering
  \begin{tabular}[c]{ccc}
    \hline\hline
    $R'$ & $\vec{N}_{R'}(\mathbf{k})$ & $R'(T_d)$ \\
    \hline
    $A_{1u}$ & $\frac{2}{\sqrt{5}}\mathbf{k}\cdot\mathbf{J}$ & $A_{2}$\\
    $T_{1u}$ & $\frac{\sqrt{6}}{\sqrt{5}}\mathbf{J}\times\mathbf{k}$ & $T_{2}$ \\
    $E_u$ & $\frac{\sqrt{2}}{\sqrt{5}}(-J_x k_x-J_y k_y+2J_z
            k_z,\sqrt{3}(J_x k_x-J_y k_y))$ & $E$\\
    $T_{2u}$ & $\frac{\sqrt{6}}{\sqrt{5}}(J_y k_z+J_z k_y,J_x k_z+J_zk_x,J_x k_y+J_y k_x)$ & $T_1$ \\
    \hline
    $T_{2u}$ & $-2\Gamma_{45}\mathbf{k}$ & $T_1$\\
    $A_{1u}$ &
               ${\boldsymbol{\mathcal{J}}}\cdot\mathbf{k}$ & $A_2$\\
    $E_u$ &
            $\frac{1}{\sqrt{2}}(-k_x\mathcal{J}_x-k_y\mathcal{J}_y+2k_z\mathcal{J}_z,\sqrt{3}(k_x\mathcal{J}_x-k_y\mathcal{J}_y))$ & $E$\\
    $T_{1u}$ &
               $\frac{\sqrt{3}}{\sqrt{2}}{\boldsymbol{\mathcal{J}}}\times\mathbf{k}$
                                & $T_2$\\
    $T_{2u}$ & $\frac{\sqrt{3}}{\sqrt{2}}(\mathcal{J}_y
               k_z+\mathcal{J}_z k_y,\mathcal{J}_x k_z+\mathcal{J}_z
               k_x,\mathcal{J}_x k_y+\mathcal{J}_y k_x)$ & $T_1$\\
    $A_{2u}$ & $\frac{1}{\sqrt{3}}\mathbf{T}\cdot\mathbf{k}$ & $A_1$ \\
    $E_u$ & $\frac{1}{\sqrt{6}}(T_x k_x+T_y k_y-2T_z
            k_z,\sqrt{3}(T_x k_x-T_y k_y))$ & $E$ \\
    $T_{1u}$ & $\frac{1}{\sqrt{2}}(T_y k_z+T_z k_y,T_x k_z+T_z k_x,T_x
               k_y+T_y k_x)$ & $T_2$\\
    $T_{2u}$ & $\frac{1}{\sqrt{2}}\mathbf{T}\times\mathbf{k}$ & $T_1$\\
    \hline\hline
  \end{tabular}
  \caption{Odd parity pairing matrices
    $\vec{N}_{R'}(\mathbf{k})$ with a single power of $\mathbf{k}$ in cubic
    symmetry with inversion $O_h$, and in tetrahedral symmetry $T_d$
    (read the representation labels on the right-hand-side).}
  \label{tab:tabtab}
\end{table}

\subsection{Tetrahedral symmetry $T_d$}
\label{sec:tetrahedral-symmetry}

In tetrahedral symmetry, $\mathbf{k}$
transforms according to $T_2$ (instead of $T_{1u}$ in cubic symmetry)
so that one needs only modify the symmetric pairing functions labels
$A_{1u}\rightarrow A_2$, $A_{2u}\rightarrow A_1$, $E_u\rightarrow E$,
$T_{1u}\rightarrow T_2$ and $T_{2u}\rightarrow T_1$ (see the
right-most column of Table~\ref{tab:tabtab}). The basis
matrices $M_R$ are unchanged except for the drop of the $g$ subscript.

\section{Fierz identities}
\label{sec:fierz-identities}

Fierz identities \cite{fierz1937,vafek2014,boettcher2016} are
reordering relations for four-fermion interactions: if $A$ and $B$ are
two $n\times n$ matrices, and $\psi_i$ $n$-component fermion fields,
there exist matrices $A',B',A'',B''$ such that
\begin{eqnarray}
  \label{eq:59}
(\psi_1^\dagger A\psi_2) (\psi_3^\dagger B\psi_4)&=&(\psi_1^\dagger
     A'\psi_4) (\psi_3^\dagger B'\psi_2)\\
&=&(\psi_1^\dagger A''\psi_3^\dagger) (\psi_4 B''\psi_2)\\
&=&-(\psi_1^\dagger A''\psi_3^\dagger) (\psi_2 B''{}^T\psi_4),
\end{eqnarray}
by virtue of the simple anticommutation relations between field operators.
Ultimately, these identities correspond to a change of basis for tensor
products. Here we do not derive Fierz identities in great generality,
but rather focus on special cases useful for our purposes.

\subsection{Derivation}
\label{sec:derivation}

Let $\{Q_a\}_{a=1,..,n^2}$ be an orthonormal basis of the Hilbert space of $n\times
n$ matrices. (In particular ${\rm Tr}[Q_aQ_b^\dagger]=n\,\delta_{ab}$.) Then, any matrix $A$ in that space can be expanded
following
\begin{equation}
  \label{eq:42}
  A=\sum_a A^a Q_a,\qquad\mbox{where}\qquad A^a=\frac{1}{n}{\rm
    Tr}[A^\dagger Q_a].
\end{equation}
A set of basis matrices can be chosen as basis matrices of the
irreducible representations of the symmetry group forming the Hilbert
space. We call such a set $\{\vec{W}_R\}_R$, where the dimension of
each vector $\vec{W}_R$ is that of the dimension of $R$. We take ${\rm Tr}[W_R^iW_{R'}^j{}^\dagger]=n\,\delta_{ij}\,\delta_{RR'}$. 

\subsubsection{Particle-hole relation}
\label{sec:part-hole-relat}

Elements of the trivial representations can be formed out of every
representation as follows:
\begin{equation}
  \label{eq:57}
  \vec{W}_R\cdot\vec{W}_R^\dagger \equiv \sum_{i=1}^{{\rm dim}R} W_R^i\otimes W_R^i.
\end{equation}
For a given representation $R_o$, we wish to find the coefficients
$f(R_o,R)$ such that
\begin{equation}
[\vec{W}_{R_o}]_{\alpha\beta}\cdot[\vec{W}_{R_o}^\dagger]_{\mu\nu} = \sum_R f(R_o,R) [\vec{W}_{R}]_{\alpha\nu}\cdot[\vec{W}_{R}^\dagger]_{\mu\beta}.
  \label{eq:58a}
\end{equation}
Multiplying Eq.~\eqref{eq:58a} by $W^i_{R_1,\lambda\alpha}{}^\dagger
W^i_{R_1,\rho\mu}$ and summing over $\alpha$ and $\mu$, we find
\begin{eqnarray}
  \label{eq:58}
 &&
    \sum_{j=1}^{{\rm dim}R_o}[W_{R_1}^i{}^\dagger W^j_{R_o}]_{\lambda\beta}[W^i_{R_1}W^j_{R_o}{}^\dagger]_{\rho\nu}\\
&&\qquad\qquad = \sum_R f(R_o,R)\sum_{j=1}^{{\rm dim}R} [W_{R_1}^i{}^\dagger W^j_{R}]_{\lambda\nu}[W_{R_1}^iW^j_{R}{}^\dagger]_{\rho\beta}.\nonumber
\end{eqnarray}
Now taking $\lambda=\nu$ and $\rho=\beta$ and summing over
$\lambda,\rho$, we find:
\begin{equation}
  \label{eq:7}
  f(R_o,R_1)=\frac{1}{n^2}\sum_{j=1}^{{\rm dim}R_o}{\rm Tr}[W_{R_1}^i{}^\dagger W^j_{R_o}W^i_{R_1}W^j_{R_o}{}^\dagger]
\end{equation}
for any $i=1,..,{\rm dim} R_o$.

\subsubsection{Particle-particle relation}
\label{sec:part-part-relat}

Similarly, we wish to find the coefficients $g(R_o,R)$ such that
\begin{equation}
  \label{eq:9}
  [\vec{W}_{R_o}]_{\alpha\beta}\cdot[\vec{W}_{R_o}^\dagger]_{\mu\nu} = \sum_R g(R_o,R) [\vec{W}_{R}\Lambda]_{\alpha\mu}\cdot[\Lambda^T\vec{W}_{R}^\dagger]_{\nu\beta},
\end{equation}
where here we have $\overline{R}=R\Lambda$, with $\Lambda^T=-\Lambda$,
$\Lambda^T\Lambda=\Lambda\Lambda^T={\rm Id}_n$. Here, we multiply Eq.~\eqref{eq:9}
by $[\Lambda^TW^i_{R_1}{}^\dagger]_{\lambda\alpha}
[W^i_{R_1}\Lambda]_{\rho\nu}$ and sum over $\alpha,\nu$:
\begin{eqnarray}
  \label{eq:35}
  &&
    \sum_{j=1}^{{\rm dim}R_o}[\Lambda^TW_{R_1}^i{}^\dagger
     W^j_{R_o}]_{\lambda\beta}[W^i_{R_1}\Lambda W^j_{R_o}{}^*]_{\rho\mu}\\
&&\quad\quad = \sum_R g(R_o,R)\sum_{j=1}^{{\rm dim}R} [\Lambda^TW_{R_1}^i{}^\dagger W^j_{R}\Lambda]_{\lambda\mu}[W_{R_1}^iW^j_{R}{}^\dagger]_{\rho\beta},\nonumber
\end{eqnarray}
and we obtain, setting $\lambda=\mu$ and $\rho=\beta$ and summing over
$\lambda,\rho$:
\begin{eqnarray}
  \label{eq:41}
   g(R_o,R_1)&=&\frac{\eta_{R_o}}{n^2}\sum_{j=1}^{{\rm
  dim}R_o}{\rm Tr}[W_{R_1}^i{}^\dagger
  W^j_{R_o}W^i_{R_1}W^j_{R_o}{}^\dagger]\nonumber\\
&=&\eta_{R_o}f(R_o,R_1),
\end{eqnarray}
where $\eta_R=\pm1$ is such that $\Lambda W^j_R{}^*\Lambda^T=\eta_R W^j_R{}^\dagger$.

\section{Eliashberg theory}
\label{sec:eliashberg-theory}

\subsection{Details of calculations from the main text}
\label{sec:deta-calc-from}

\subsubsection{Spherical (or cubic) harmonic decomposition}
\label{sec:spherical-or-cubic}

The components of the interaction $V_{0,1}$ defined in Eq.~\eqref{eq:38} are
\begin{eqnarray}
  \label{eq:60}
  V_0(|\mathbf{k}|,|\mathbf{k}'|;\w-\w')&=&\frac{1}{4\pi}\int
                                                        d\mathbf{\hat{k}}d\mathbf{\hat{k}}'V(\mathbf{k}-\mathbf{k}',\w-\w')\\
V_1(|\mathbf{k}|,|\mathbf{k}'|;\w-\w')&=&\frac{1}{4\pi}\int d\mathbf{\hat{k}}d\mathbf{\hat{k}}'(\mathbf{\hat{k}}\cdot\mathbf{\hat{k}}')V(\mathbf{k}-\mathbf{k}',\w-\w'),\nonumber
\end{eqnarray}
where $\int d\mathbf{\hat{k}}=\int_0^\pi d\theta
\sin\theta\int_{0}^{2\pi}d\phi$. 

\subsubsection{Projected representation mixing}
\label{sec:proj-repr-mixing}

When several copies of a representation appear, one must solve for a
mixture of matrices belonging to each copy. In the $J=2$ case, defining
\begin{equation}
  \label{eq:21}
  \Sigma_A(k)=\Delta(\omega)\sum_{i=1,2}\phi_in^\alpha_{2,i}(\mathbf{\hat{k}}),
\end{equation}
one must now solve
\begin{equation}
  \label{eq:5}
\sum_{i,i'}n^\alpha_{J,i}(\mathbf{\hat{k}})\left[\Delta(\omega)\delta_{ii'}+\frac{\pi}{\beta_c}\mathcal{L}_J(\omega)A_J^{ii'}\right]\phi_{i'}=0,
\end{equation}
where 
\begin{equation}
  \label{eq:10}
  \mathcal{L}_j(\omega)=\sum_{\omega'}\frac{\Delta(\omega')}{|\omega'|}f_J(\omega-\omega').
\end{equation}
This is equivalent to solving
\begin{equation}
  \label{eq:11}
  \mathbf{A}_J(\Delta(\omega)+\frac{\pi}{\beta_c}\mathcal{L}_J(\omega)\mathbf{A}_J)=\mathbf{0},
\end{equation}
where $\mathbf{A}_J=(A_J^{ii'})_{ii'}$ (the fifth row of
Table~\ref{tab:strengths}), and hence a set of solutions is given by
solving for the eigenvalues and eigenvectors of $\mathbf{A}_J$.

The eigenvalues and eigenvectors of $A_{J=2,ii'}$ of
Table~\ref{tab:strengths} lead to the following pairing strengths and
matrices. For the $3/2$ bands:
\begin{equation}
  \label{eq:32}
\begin{cases}
\tilde{A} = 27/70,&
  \tilde{n}_{J=2,1},=\frac{1}{\sqrt{15}}(-\sqrt{14}n_{J=2,S=1}+n_{J=2,S=3})\\
\tilde{A} = 0,&
  \tilde{n}_{J=2,2}=\frac{1}{\sqrt{15}}(n_{J=2,S=1}+\sqrt{14}n_{J=2,S=3})
\end{cases},
\end{equation}
and for the $1/2$ bands:
\begin{equation}
  \label{eq:33}
\begin{cases}
\tilde{A} =\frac{55+\sqrt{2689}}{140},&
  \tilde{n}_{J=2,1}=\sqrt{\frac{1}{2}-\frac{79}{10\sqrt{2689}}}n_{J=2,S=1}\\
 &\qquad\qquad+\sqrt{\frac{1}{2}+\frac{79}{10\sqrt{2689}}}n_{J=2,S=3}\\
\tilde{A} = \frac{55-\sqrt{2689}}{140},&
  \tilde{n}_{J=2,2}=-\sqrt{\frac{1}{2}+\frac{79}{10\sqrt{2689}}}n_{J=2,S=1}\\
&\qquad\qquad+\sqrt{\frac{1}{2}-\frac{79}{10\sqrt{2689}}}n_{J=2,S=3}
\end{cases}.
\end{equation}

\end{document}